\documentclass[preprint,12pt]{elsarticle}
\usepackage{amsmath}
\usepackage{hyperref}

\usepackage{amssymb}

\journal{Icarus}

\begin{document}

\begin{frontmatter}



\title{Doubly Synchronous Binary Asteroid Mass Parameter Observability\tnoteref{t1}}
\tnotetext[t1]{This material is based upon work supported by the National Science Foundation Graduate Research Fellowship Program under Grant No. DGE 1650115. Any opinions, findings, and conclusions or recommendations expressed in this material are those of the author(s) and do not necessarily reflect the views of the National Science Foundation.}


\author[1]{Alex B. Davis\corref{cor1}%
}
\ead{Alex.B.Davis@Colorado.edu}
\author[1]{Daniel J. Scheeres}
\ead{Scheeres@Colorado.edu}

\cortext[cor1]{Corresponding author}
\address[1]{Department of Aerospace Engineering Science, University of Colorado Boulder, CO 80309, United States}

\begin{abstract}
The full two-body problem (F2BP) is often used to model binary asteroid systems, representing the bodies as two finite mass distributions whose dynamics are influenced by their mutual gravity potential. The emergent behavior of the F2BP is highly coupled translational and rotational mutual motion of the mass distributions. A large fraction of characterized binary asteroids appear to be at, or near, the doubly synchronous equilibrium, which occurs when both bodies are tidally-locked and in a circular co-orbit. Stable oscillations about this equilibrium can be shown, for the nonplanar system, to be combinations of seven fundamental frequencies of the system and the mutual orbit rate. The fundamental frequencies arise as the linear periods of center manifolds identified about the equilibrium which are heavily influenced by each body's mass parameters. We leverage these eight dynamical constraints to investigate the observability of binary asteroid mass parameters via dynamical observations. This is accomplished by deriving a relationship between the fundamental frequencies and mass parameters for doubly synchronous systems. This relationship allows us to show the sensitivity of the dynamics to changes in the mass parameters, first for the planar dynamics, and then for the nonplanar dynamics. In so doing we are able to predict the idealized estimation covariance of the mass parameters based on observation quality and define idealized observation accuracies for desired mass parameter certainties. We apply these tools to 617 Patroclus, a doubly synchronous Trojan binary and flyby target of the LUCY mission, as well as the Pluto and Charon system in order to predict mutual behaviors of these doubly synchronous systems and to provide observational requirements for these systems mass parameters.
\end{abstract}




\end{frontmatter}


\section{Introduction}
\indent In the past three decades binary asteroids have been discovered throughout the solar system and are believed to make up a significant portion of many small body populations. Observers have identified 48 transneptunian binaries and place binaries at approximately 16$\%$ of near-Earth asteroids (NEAs); the majority of equal mass binaries in these populations are explected to be doubly synchronous\cite{noll2007binaries}\cite{margot2002binary}\cite{pravec2006photometric}. Radar observations of Near Earth Asteroids (NEAs), such as 1999 KW4 and 65803 Didymos, have generated shape models for a number of these bodies. Such information provides a basis for the application of the full two-body problem (F2BP) to study the behavior of these systems\cite{fahnestock2006simulation}\cite{naidu2015near}. The F2BP describes the dynamical interactions of two mass distributions, which result in coupled translational and rotational motion due to the mutual gravity potential between asymmetric mass distributions. The dynamics, stability and the effects of mass and spin for the F2BP have been studied extensively by Maciejewski, Tricario, Scheeres, and Boue and Laskar amongst others\cite{maciejewski1995reduction}\cite{tricarico2008figure}\cite{scheeres2009stability}\cite{boue2009spin}. Work by Fahnestock and Scheeres implemented a polyhedral formulation of the F2BP dynamics to simulate the behavior of 1999 KW4 based on a series of radar observations\cite{fahnestock2006simulation}. Later work by Naidu adapted the inertia integral implementation of the F2BP developed by Ashenberg to study the spin behavior of a sampling of 10 well-observed binary NEAs\cite{naidu2015near}\cite{ashenberg2007mutual}. Both studies provided valuable insight into the dynamical behavior of binaries, however they were limited by the computational burden or limited expansion order inherent to their implementation of the F2BP. Recently Hou et al. derived a recursive approach to the F2BP which enables much more computationally efficient simulation of the F2BP\cite{hou2016mutual}. The improvements in computationally efficiency also open the door to study mass distribution sensitivity of binary system dynamics.\\
\indent As observers continue to study binary systems in more detail it is important to understand how their mass parameters may influence the observed dynamics and affect assumptions made about the system behavior. We analyze these effects by applying estimation techniques based on idealized observations of the translational and rotational coupling inherent in a doubly synchronous binary system. Such an approach, while unprecedented for binary asteroids, has been leveraged during missions to small bodies and asteroid flybys of the Earth. Most notably, Takahashi and Scheeres used observations of the spin state of 4179 Toutatis during several Earth flybys to estimate its moments of inertia\cite{takahashi2013spin}. At Vesta the Dawn mission was able to use the spacecraft's orbital behavior to place constraints on the interior structure and mass distribution of Vesta and similar plans have been made for the OSIRIS-Rex mission to Bennu\cite{ermakov2014constraints}\cite{scheeres2016geophysical}. Development of these capabilities for binaries would enable more reliable measurements and more robust mission planning for upcoming missions to binary asteroids, such as the LUCY mission which will fly by the 617 Patroclus binary system and the AIDA mission which will impact and observe 65803 Didymos.\\
\indent In this paper we leverage a new formulation of the F2BP dynamics developed in Hou et al. to explore the ideal observability of binary asteroids mass parameters from observations of the dynamics. The Hou formulation of the F2BP is selected because its recursive form provides improved computation speed and ease of manipulation\cite{hou2016mutual}. To understand the sensitivity of the binary systems to mass parameters we study the observability of each mass parameter. To maintain simplicity of the investigation the simulated binary system is assumed to be in or oscillating near a doubly synchronous equilibrium, meaning both bodies are near tidally locked in a relaxed orbit about their mutual center of mass.  We also assume knowledge of the center of mass and principal axes of the target bodies so as to better study the direct effects of the mass parameters themselves on the observable dynamics. Thus, this paper provides ideal limits on the observability of the mass parameters based purely on observations of a binary system's mutual motion.\\
\indent Our analysis is primarily applied to the 617 Patroclus system, a Trojan asteroid believed to be a near equal mass doubly synchronous binary system made up of two nearly ellipsoidal bodies. 617 Patroclus thus represents a realistic manifestation of the doubly synchronous assumptions made in this dynamical analysis. Additionally as a flyby target for the LUCY mission, this binary has been the target of a number of recent observation campaigns. The physical parameters describing 617 Patroclus implemented for this study are listed in Table 1 and based on the results of the 2013 stellar occultation study performed by Buie et al. In line with the analysis of Buie et al. we assume the system to be homogenous of constant density\cite{buie2015size}. Additionally, we apply this technique to the Pluto-Charon system, exploring its applicability to near-spherical systems.

In this paper, we will first rigorously define the F2BP, the mutual gravity potential and the doubly synchronous equilibrium. With this information we explore the dynamics of the planar form of the F2BP; analyzing the fundamental frequencies of the planar dynamics and their observability. This analysis is then expanded to the nonplanar form of the dynamics and an idealized mass parameter estimation method is developed and explored. Finally, we examine the case of near-spherical binaries by applying our methods to the Pluto-Charon system.

Throughout the paper we refer to observations of a binary system's fundamental frequencies, which we assume come from an idealized fictitious observer. The fictitious observer is imagined to receive regular resolved images of the system of interest, such that the dynamics associated with each fundamental frequency can be explicitly measured. The reasons for this approach are twofold. Firstly, this approach allows us to perform a best-case analysis and provide an upper bound of what is achievable with the described methodology. Secondly, it allows us to avoid systemic and stochastic errors from a given measurement type, such as the effect of uncertainty in shape and surface properties on light curve data, which we view to be outside the scope of this work.

\begin{table}[]
\centering
\caption{Physical Properties of The Doubly Synchronous 617 Patroclus modeled in this study}
\begin{tabular}{| c | c | c | c | c | c | c | c | c | c |}
\hline
\multicolumn{1}{|c|}{\textbf{Density}}&\multicolumn{3}{|c|}{\textbf{Primary}}&\multicolumn{3}{|c|}{\textbf{Secondary}}&\multicolumn{2}{|c|}{\textbf{Orbit}}\\\hline
\textbf{$\rho$ [kg/m$^{-3}$]} & a [km] & b [km] & c [km] &a [km]&b [km] & c [km] & r [km] & Period [days]\\\hline
881&63.5&58.5&49&58.5&54&45&664.6&4.41\\\hline
\end{tabular}
\end{table}

\section{The Full Two-Body Problem}

In order to model the dynamics of binary asteroids we implement a simulation of the F2BP. The F2BP describes the mutual gravitational interactions between two arbitrary mass distributions, as illustrated in Fig. 1. To model the mutual gravity of the mass distributions the influence of each infinitesimal mass element of both bodies on all other mass elements in the opposing body must be accounted for. The self gravitational potential is ignored in this implementation of the F2BP as both bodies are assumed to be rigid for the time scales of interest and thus their self potentials are constant. The generic form of the mutual gravity potential is a double volume integral over both mass distributions, where $d$ is the the distance between an finite mass element of the primary,  $dm_A$, and the secondary, $dm_B$. 

\begin{gather}
U=-G\int_A\int_B \frac{1}{d}dm_A dm_B
\end{gather}
\begin{figure}[!htb]
    \centering
    \includegraphics[width=.75\textwidth]{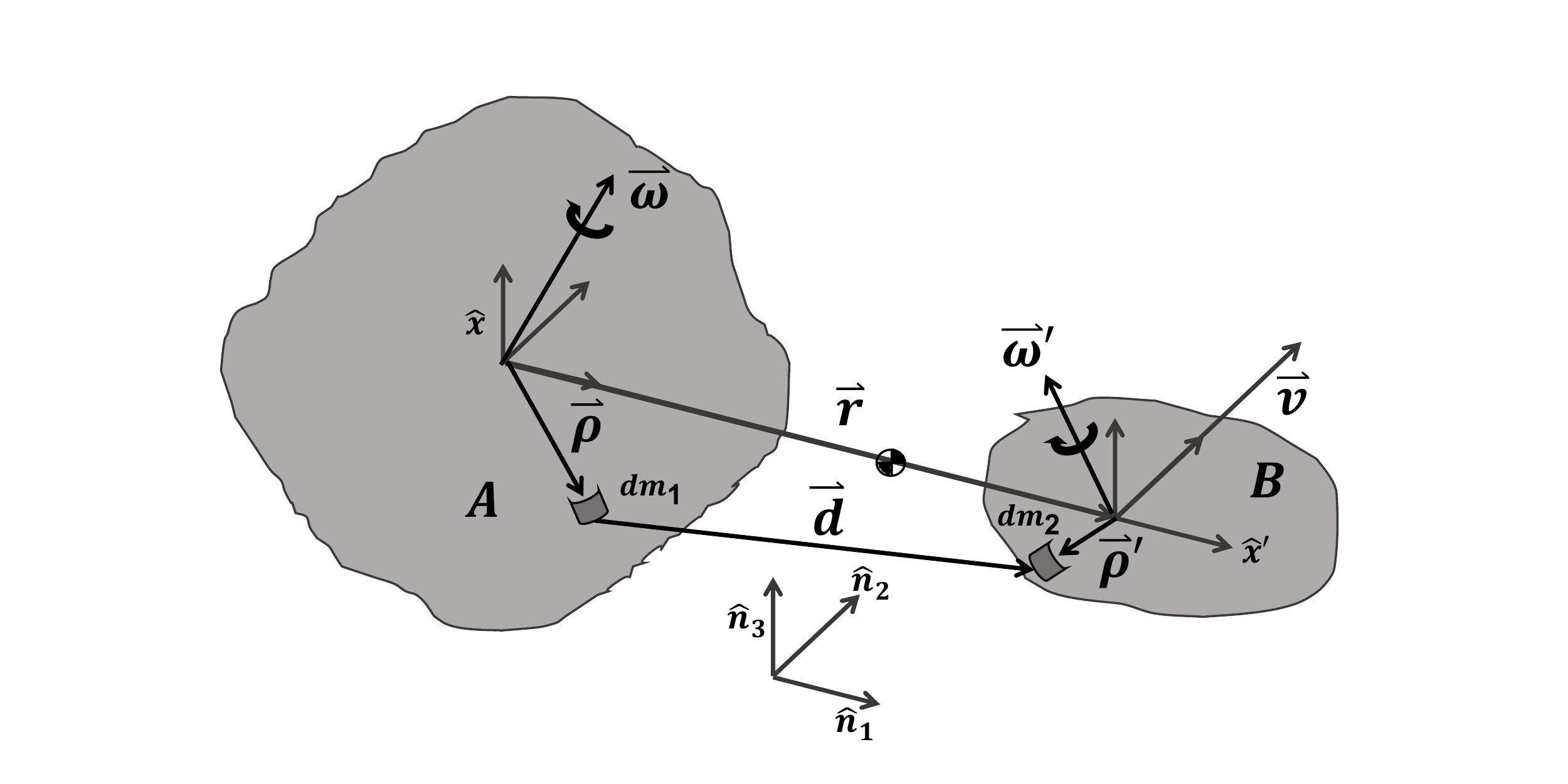}
    \caption{Diagram of initial system geometry. $\hat{x}$ represents the principal frame of the primary, $A$, $\hat{x}'$ is the principal frame of the secondary, $B$, and $\hat{n}$ is the inertial frame. The finite mass elements $dm_A$ and $dm_B$ must be integrated over the full body using their locations relative to the body centers of mass, $\vec{\rho}$ and $\vec{\rho'}$, to compute the mass parameters and mutual gravity potential.}
\end{figure}

In the Hou et al. reformulation the mutual gravity potential is expressed as a function of the relative separation magnitude, $R$, and the attitude of the two approximately modeled asteroids, $\textbf{\textit{C}}_A$ and $\textbf{\textit{C}}_B$; where the attitude matrices represent a mapping from the indicated body's principal frame into the inertial frame.  At the second degree and order of the mass distribution, the mutual potential is:

\begin{gather}
U=-\frac{GM_A M_B}{R}-\frac{G}{2R^3}\Big(M_B \mbox{Tr}(\textbf{\textit{I}}_A)+M_A \mbox{Tr}(\textbf{\textit{I}}_B)\Big)\\
\nonumber+ \frac{3G}{2R^5}\vec{r}\bullet\Big(M_B\textbf{\textit{C}}_A\textbf{\textit{I}}_A\textbf{\textit{C}}_A^T+M_A \textbf{\textit{C}}_B\textbf{\textit{I}}_B\textbf{\textit{C}}_B^T\Big)\bullet\vec{r}
\end{gather}

The second order mass distribution terms are the principally aligned inertia tensors, $\textbf{\textit{I}}_A$ and $\textbf{\textit{I}}_B$, expressed as linear combinations of the inertia integrals:

\begin{gather}
\textbf{\textit{I}}_i=M_i\begin{bmatrix}
T^{0,2,0}_i +T^{0,0,2}_i&0&0\\
0&T^{2,0,0}_i +T^{0,0,2}_i&\\
0&0&T^{2,0,0}_i +T^{0,2,0}_i\\
\end{bmatrix}
\end{gather}

The inertia integrals capture the mass distributions up to a truncation order N, here selected as the second order. In their generic form the inertia integrals are analogous to spherical harmonics while more similar in structure to the moments and products of inertia\cite{tricarico2008figure}. 

\begin{gather}
T^{l,m,n}=\frac{1}{MR^{l+m+n}}\int_B x^l y^m z^n dm \mbox{, where }l+m+n=N
\end{gather}

A more general form of the mutual gravity potential may be found in the 2016 Hou et al. paper.

\subsection{Equations of Motion}
Given the form of the mutual gravity potential the equations of motion (EOMs) for the F2BP can be generated. The inertial form of the F2BP has 12 degrees of freedom, however, using the relative dynamics it can be reduced to 9 degrees of freedom\cite{maciejewski1995reduction}.
\begin{gather}
\vec{X}=\begin{bmatrix}
\vec{r}& \vec{\theta}_1& \vec{\theta}_2& \dot{\vec{r}}& \vec{\omega}_1& \vec{\omega}_2
\end{bmatrix}^T
\end{gather}
where $\vec{r}$ is the relative separation vector measured from the primary to secondary, the vectors $\vec{\theta}_1$ and $\vec{\theta}_2$ are Euler 123 angle sets defining the inertial orientation of the primary and the orientation of the secondary relative to the primary, the vectors $\vec{\omega}_1$ and $\vec{\omega}_2$ are the angular velocities corresponding to the Euler angles such that

\begin{gather}
\dot{\vec{\theta}}_i=\textbf{\textit{B}}_{i}\vec{\omega}_i\\
\textbf{\textit{B}}_{i}=\frac{1}{\cos\theta_{i,2}}\begin{bmatrix}
\cos\theta_{i,3} & -\sin\theta_{i,3} & 0\\
\cos\theta_{i,2} \sin\theta_{i,3} & \cos\theta_{i,2} \cos\theta_{i,3} &0\\
-\sin\theta_{i,2} \cos\theta_{i,3} &\sin\theta_{i,2} \sin\theta_{i,3} &\cos\theta_{i,2}
\end{bmatrix}
\end{gather}

The rotation matrix corresponding to the Euler angle sets are 

\begin{gather}
\textbf{\textit{C}}_i=\textbf{\textit{C}}(\vec{\theta}_i)
\end{gather}

where the Euler angle set is the standard Euler 123 body set. Although non-singular attitude representations are used to numerically integrate the binary system, we use Euler angles to describe the attitude because they have well-behaved and intuitive linearization properties for analysis later in this paper.

The rotation matrices corresponding to the two Euler angle sets, $\vec{\theta}_1$ and $\vec{\theta}_2$, are then

\begin{gather}
\textbf{\textit{C}}_A=\textbf{\textit{C}}(\vec{\theta}_1)\\
\textbf{\textit{C}}_{B/A}=\textbf{\textit{C}}(\vec{\theta}_2)=\textbf{\textit{C}}_A^T\textbf{\textit{C}}_B
\end{gather}

Where $\vec{\theta}_2$ is selected as the mapping from the secondary's frame into the primary's frame for convenience of later analysis.

Following the implementation of Maciejewski, the F2BP EOMs can be derived from Newton's second law and Euler's EOMs as \cite{maciejewski1995reduction}.  
\begin{gather}
\ddot{\vec{r}}=- \textbf{\textit{I}}_A^{-1} \Big(  \textbf{\textit{I}}_A \vec{\omega}_1 \tilde{\omega}_1 + \vec{M}_A \Big)\tilde{r} - 2\tilde{\omega}_1 \dot{\vec{r}} - \tilde{\omega}_1 \tilde{\omega}_1 \vec{r} - \frac{1}{m}\frac{\partial U}{\partial \vec{r}}\\
\dot{\vec{\omega}}_1 =  \textbf{\textit{I}}_A^{-1} \Big(  \textbf{\textit{I}}_A \vec{\omega}_1 \tilde{\omega}_1 + \vec{M}_A \Big)\\
\dot{\vec{\omega}}_2 = \textbf{\textit{C}}_{B/A} \textbf{\textit{I}}_B^{-1} \textbf{\textit{C}}_{B/A}^T\Big( \textbf{\textit{C}}_{B/A} \textbf{\textit{I}}_B \textbf{\textit{C}}_{B/A}^T\Big( \vec{\omega}_1+\vec{\omega}_2\Big) \tilde{\omega}_1 +\vec{M}_B\\
\nonumber+ \Big(\dot{\textbf{\textit{C}}}_{B/A}\textbf{\textit{I}}_B \textbf{\textit{C}}_{B/A}^T +\textbf{\textit{C}}_{B/A} \textbf{\textit{I}}_B \dot{\textbf{\textit{C}}}_{B/A}\Big)\cdot \Big( \vec{\omega}_1+\vec{\omega}_2\Big)\Big)\\
\nonumber - \textbf{\textit{I}}_A^{-1} \Big( \textbf{\textit{I}}_A \vec{\omega}_1 \tilde{\omega}_1 + \vec{M}_A \Big)
\end{gather}

Where the $\tilde{(-)}$ operator describes the skew-symmetric matrix transform on a vector in $\mathbb{R}^3$

\begin{gather}
\tilde{f}=\begin{bmatrix}
0&-f_3&f_2\\
f_3&0&-f_1\\
-f_2&f_1&0
\end{bmatrix}
\end{gather}

To fully implement the F2BP EOMs the mutual gravitational torques, $\vec{M_A}$ and $\vec{M_B}$, must be accounted for, this is accomplished by taking the partials of the mutual gravity with respect to the relative states of the system.
\begin{gather}
\vec{M_B}=-\vec{\alpha}\times\dfrac{\partial U}{\partial \vec{\alpha}}-\vec{\beta}\times\dfrac{\partial U}{\partial \vec{\beta}}-\vec{\gamma}\times\dfrac{\partial U}{\partial \vec{\gamma}}
\end{gather}
\begin{gather}
\vec{M}_A=\vec{r}\times\dfrac{\partial U}{\partial \vec{R}}-\vec{M}_B
\end{gather}
Where $\vec{\alpha}$, $\vec{\beta}$, and $\vec{\gamma}$ are columns of the rotation matrix
\begin{gather}
\textbf{\textit{C}}_{B/A}=\begin{bmatrix}
\vec{\alpha}&\vec{\beta}&\vec{\gamma}
\end{bmatrix}
\end{gather}

The derivation of this formulation of the gravity torques is further described in Maciejewski\cite{maciejewski1995reduction}.

We choose not to include the effects of external perturbers, such as the Sun and planets, because their influence on the relative dynamics is small and would be mostly averaged out of the dynamics as the asteroids repeat their orbit.

\subsection{Doubly Synchronous Equilibrium}
The high dimensionality of the nonplanar form of the F2BP makes dynamical analysis of the general system complex and unwieldy; we thus turn to the doubly synchronous equilibrium for use as a simplified dynamical sandbox. For the F2BP the doubly synchronous equilibrium occurs when the bodies are aligned with their long axes facing each other, tidally locked and co-orbiting. Two such equilibria exist in the F2BP, an inner and outer solution, however only the outer solution is stable and seen in nature\cite{scheeres2009stability}. For the remainder of this paper only the outer stable equilibrium will be referred to as the doubly synchronous equilibrium. 
In order to compute the doubly synchronous equilibrium for a given system the amended potential and its partials with respect to the degrees of freedom must be analyzed\cite{scheeres2012minimum}. The amended potential is defined as
 \begin{gather}
\mathcal{E} = \frac{H^2}{2I_H}+U
\end{gather}
Where $H$ is the angular momentum magnitude and $\textbf{\textit{I}}_H$ is the moment of inertia about the angular momentum axis,
\begin{gather}
I_H=\hat{H} \bullet \Big( \textbf{\textit{I}}_A+ \textbf{\textit{I}}_B  + m\Big[R^2 \textbf{\textit{I}}_{3x3}-\vec{r}\vec{r}\Big]\Big) \bullet \hat{H}
\end{gather}
the matrix $\bar{\bar{\textbf{\textit{U}}}}$ denotes the identity matrix and $\bullet$ denotes the dot product.
The scalar $m$ is the reduced mass
\begin{gather}
m=\frac{M_AM_B}{M_A+M_B}
\end{gather}
Where $\hat{H}$ defines the unit direction of the angular momentum .
\begin{gather}
\hat{H}=\begin{bmatrix}
\cos{\delta}\\
\sin{\delta}\sin{\lambda}\\
\sin{\delta}\cos{\lambda}
\end{bmatrix}
\end{gather}
Where $\delta$ is an offset angle from the x-axis and $\lambda$ acts as a clocking angle about the x-axis. In the equilibrium state angle $\delta$ is $\pi/2$ radians and angle $\lambda$ is 0.

Based on knowledge of the doubly synchronous equilibrium orientation for a second degree and order gravity field, we can simplify $I_H$ to
\begin{gather}
I_H=I_{A,z}+I_{B,z}+mr^2
\end{gather}
The subscript $z$ denotes the z-axis moment of inertia, or the polar moment of inertia for the body.

For a given angular momentum of the system and set of mass parameters for the system, the zeroes of the partials of the amended potential with respect to the system degrees of freedom can be used to identify an equilibrium. 
\begin{gather}
\mathcal{E}_{\vec{r}},\mathcal{E}_{\dot{\vec{r}}},\mathcal{E}_{\vec{\theta}},\mathcal{E}_{\dot{\vec{\theta}}},\mathcal{E}_{\delta},\mathcal{E}_{\lambda}=0
\end{gather}
The detailed formulation of the amended potential partials used to find the extrema are
\begin{gather}
\frac{\partial \mathcal{E}}{\partial \vec{r}} = -\frac{H^2}{I_H^2}m\Big[I_{3x3}-\hat{H}\hat{H}\Big]\bullet \vec{r}+\frac{\partial U}{\partial \vec{r}}
\end{gather}
\begin{gather}
\frac{\partial \mathcal{E}}{\partial \vec{\theta}} = -\frac{H^2}{2I_H^2}\frac{\partial I_H}{\partial \vec{\theta}}+\frac{\partial U}{\partial \vec{\theta}}
\end{gather}
\begin{gather}
\frac{\partial \mathcal{E}}{\partial \xi} = -\frac{H^2}{2I_H^2}\Big[2\hat{H} \bullet \Big( \textbf{\textit{I}}_A+ \textbf{\textit{I}}_B  + m\Big[R^2 \textbf{\textit{I}}_{3x3}-\vec{r}\vec{r}\Big]\Big) \bullet \frac{\partial \hat{H}}{\partial \xi}\Big]\\
\nonumber \text{Where $\xi$ is either $\lambda$ or $\delta$}
\end{gather}

To be stable, the Hessian of the amended potential evaluated at the equilibrium point must be positive definite. 
\begin{gather}
||\mathcal{E}_{\vec{X}\vec{X}}||_{eq} > 0
\end{gather}
Through this approach it can be shown that only the outer equilibrium point is energetically stable\cite{scheeres2009stability}. We thus conclude that the outer equilibrium is the only equilibria of interest when observing natural systems and focus on this configuration as the doubly synchronous equilibrium of interest.

\section{Planar Dynamics Analysis}
To understand the system behavior at a fundamental level, we first simplify it further by applying a planar and second-order assumption. These assumptions reduce the system to four degrees of freedom, as illustrated in Fig. 2, such that it can be described as an inertial orbit angle $\theta$, a relative separation magnitude $R$, and the phase angle of each body relative to the separation line $\phi_1$ and $\phi_2$.

\begin{figure}[!htb]
    \centering
    \includegraphics[width=.75\textwidth]{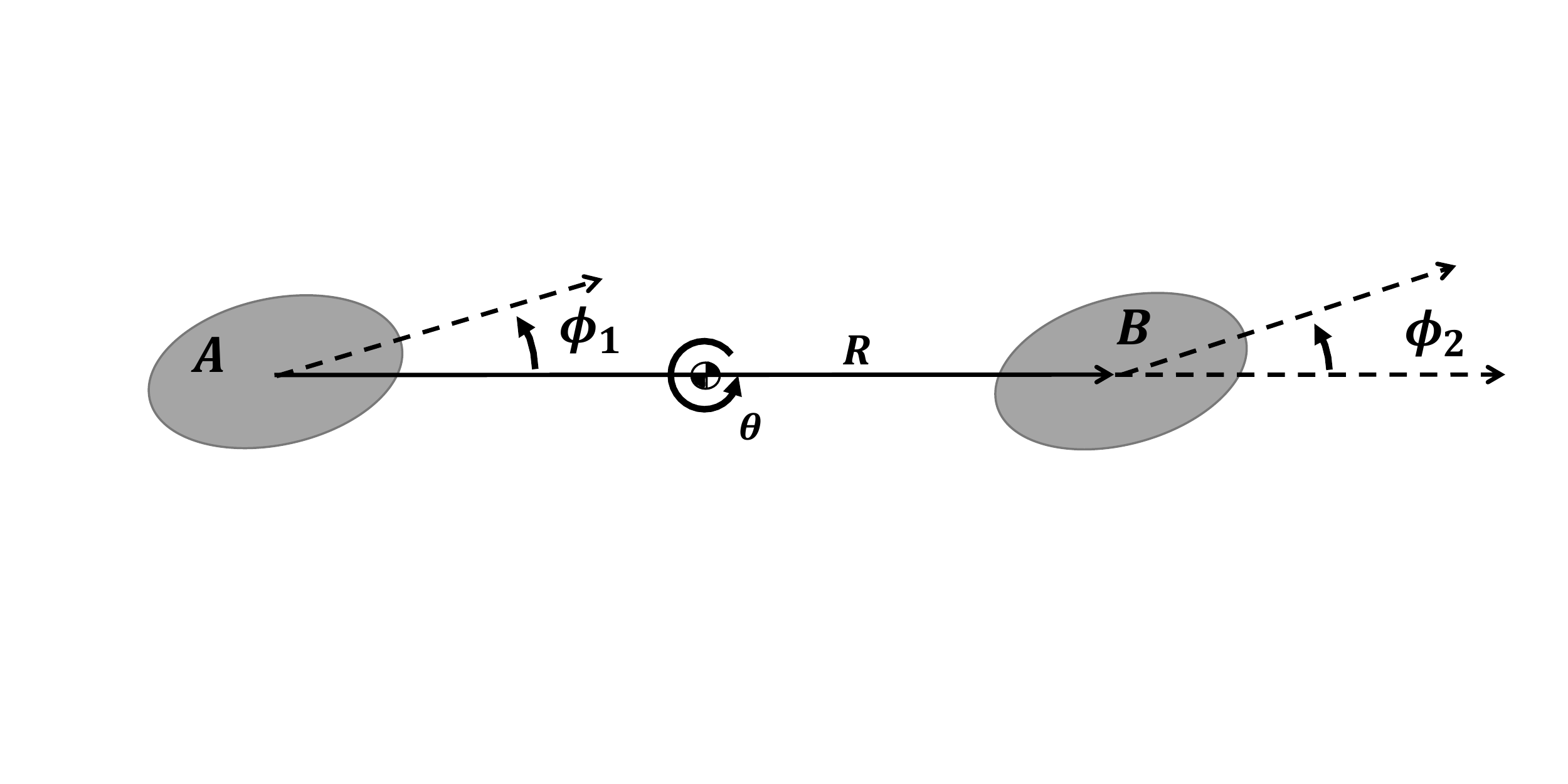}
    \caption{Diagram of the planar and second-order F2BP. The system states are an inertial orbit angle $\theta$, a relative separation magnitude $R$, and the phase angle of each body relative to the separation line $\phi_1$ and $\phi_2$.}
\end{figure}

The equations of motion for this simplified system can be derived from a Lagrangian analysis of the dynamics\cite{scheeres2009stability}. The Lagrangian for the planar problem is 

\begin{gather}
L=\frac{1}{2}I_{A, zz}\dot{\phi}_1^2+\frac{1}{2}I_{B, zz}\dot{\phi}_2^2+\frac{1}{2}m\dot{R}^2+\frac{1}{2}(I_{A, zz}+I_{B, zz}+mR^2)\dot{\theta}^2\\
\nonumber+(I_{A, zz}\dot{\phi}_1 +I_{B, zz}\dot{\phi}_2)\dot{\theta} -U_{pl}(R,\phi_1, \phi_2)
\end{gather}
Where we define the term $I_{i,jj}$ to be the mass normalized moment of inertia of body $i$ about axis $j$.

Application of Lagrange's equation yields the following EOMs
\begin{gather}
\ddot{R}=\dot{\theta}^2R-\frac{1}{m}\frac{\partial U_{pl}}{\partial R}\\
\ddot{\theta}=\frac{1}{mR^2}\frac{\partial U_{pl}}{\partial \phi_1} +\frac{1}{mR^2}\frac{\partial U_{pl}}{\partial \phi_2}-2\frac{\dot{R}\dot{\theta}}{R}\\
\ddot{\phi}_1=-\Big( 1+\frac{mR^2}{I_{A, zz}}\Big)\frac{1}{mR^2}\frac{\partial U_{pl}}{\partial \phi_1} -\frac{1}{mR^2}\frac{\partial U_{pl}}{\partial \phi_2}+2\frac{\dot{R}\dot{\theta}}{R}\\
\ddot{\phi}_2=-\Big( 1+\frac{mR^2}{I_{B, zz}}\Big)\frac{1}{mR^2}\frac{\partial U_{pl}}{\partial \phi_2} -\frac{1}{mR^2}\frac{\partial U_{pl}}{\partial \phi_1}+2\frac{\dot{R}\dot{\theta}}{R}
\end{gather}
Where $U_{pl}$ is the planar simplification of Eq. 2
\begin{gather}
U_{pl}=-\frac{GM_A M_B}{R} \bigg\{ 1+\frac{1}{2R^2} \bigg[\mbox{Tr}\Big(\bar{\textbf{\textit{I}}}_A \Big) + \mbox{Tr}\Big(\bar{\textbf{\textit{I}}}_B \Big)  -\frac{3}{2} \Big( I_{A,xx}+I_{A,yy}\\
\nonumber+I_{B,xx}+I_{B,yy} -\cos 2\phi_1 \Big(I_{A,yy}-I_{A,xx}\Big) -\cos 2\phi_2\Big(I_{B,yy}-I_{B,xx} \Big) \Big)\bigg] \bigg\}
\end{gather}
Where we define $\bar{\textbf{\textit{I}}}_i$ to be the mass normalized inertia tensor of body $i$

For this realization of the system, the conditions for the doubly synchronous equilibrium can be can restated using the system states to specify axial alignment and a circular mutual orbit\cite{scheeres2006relative}.

\begin{gather}
\phi_1=\phi_2= \dot{\phi}_1=\dot{\phi}_2=\dot{R}=0
\end{gather}

From inspection it is clear that $U_{\phi_i}$ under these conditions is equal to zero, such that

\begin{gather}
\ddot{\phi_1}=\ddot{\phi_2}=\ddot{\theta}=0
\end{gather}

Thus the necessary equilibrium rotation rate, $\dot{\theta}^*$, for an equilibrium separation, $R^*$ is 

\begin{gather}
\dot{\theta}^{*2}=\frac{1}{mR^*}\frac{\partial U_{pl}}{\partial R}\\
\dot{\theta}^{*2}=\frac{G(M_A+M_B)}{R^{*3}}\Big[1+\frac{3}{2R^{*2}}\Big[\mbox{Tr}\Big(\textbf{\textit{I}}_A\Big)+\mbox{Tr}\Big(\textbf{\textit{I}}_B\Big)-3I_{A,xx}-3I_{B,xx}\Big]\Big]
\end{gather}

The separation still must be selected to ensure the stability of the equilibrium as defined by the second partial of the amended potential in Eq. 27. With the planar and second-order assumption the relative separation can be reduced to a scalar value which allows us to identify the condition for stability\cite{scheeres2009stability}.
\begin{gather}
R^*>\sqrt{\frac{3(M_A I_{A,zz}+M_B I_{B,zz})}{m}}
\end{gather}

\subsection{Linear System Manifolds}
Because the doubly synchronous equilibrium is an energetically stable arrangement, oscillations about the equilibrium will move along closed cycles, referred to as center manifolds. In their linear form each center manifold has an associated and unique frequency, this results from the purely imaginary eigenvalue associated with the given center manifold\cite{gomez2001dynamics}. Because of the dynamical coupling of the F2BP and its relation to the mass parameters, analysis of oscillations about the doubly synchronous equilibrium may provide insight into the observability of the mass parameters. To begin this analysis, we first must identify the eigenvalues of the dynamical system. This is accomplished by computing the characteristic equation of the linearized system. The linearized dynamics are described by
\begin{gather}
\dot{\vec{X}}=\textbf{\textit{A}}\vec{X}=\begin{bmatrix}
0&0&0&0&1&0&0&0\\
0&0&0&0&0&1&0&0\\
0&0&0&0&0&0&1&0\\
0&0&0&0&0&0&0&1\\
\frac{\partial \ddot{R}}{\partial R} & 0 & \frac{\partial \ddot{R}}{\partial \phi_1} & \frac{\partial \ddot{R}}{\partial \phi_2} & 0 & \frac{\partial \ddot{R}}{\partial \dot{\theta}} & 0 &0\\
\frac{\partial \ddot{\theta}}{\partial R} & 0 & \frac{\partial \ddot{\theta}}{\partial \phi_1} & \frac{\partial \ddot{\theta}}{\partial \phi_2} & \frac{\partial \ddot{\theta}}{\partial \dot{R}} & \frac{\partial \ddot{\theta}}{\partial \dot{\theta}} & 0 &0\\
\frac{\partial \ddot{\phi_1}}{\partial R} & 0 & \frac{\partial \ddot{\phi_1}}{\partial \phi_1} & \frac{\partial \ddot{\phi_1}}{\partial \phi_2} & \frac{\partial \ddot{\phi_1}}{\partial \dot{R}} & \frac{\partial \ddot{\phi_1}}{\partial \dot{\theta}} & 0 &0\\
\frac{\partial \ddot{\phi_2}}{\partial R} & 0 & \frac{\partial \ddot{\phi_2}}{\partial \phi_1} & \frac{\partial \ddot{\phi_2}}{\partial \phi_2} & \frac{\partial \ddot{\phi_2}}{\partial \dot{R}} & \frac{\partial \ddot{\phi_2}}{\partial \dot{\theta}} & 0 &0
\end{bmatrix}\begin{bmatrix}
R\\\theta\\\phi_1\\\phi_2\\\dot{R}\\\dot{\theta}\\\dot{\phi_1}\\\dot{\phi_2}
\end{bmatrix}
\end{gather}

The eigenvalues of the system are roots of the characteristic equation

\begin{gather}
|\textbf{\textit{A}}-\lambda \textbf{\textit{I}}_{8x8}|=0
\end{gather}

which can be related to a period of motion of the linearized oscillations
\begin{gather}
P_{\beta_i}=\frac{2\pi}{\textbf{\textit{Im}}(\lambda_i)}
\end{gather}

Here $\beta_i$ specifies the imaginary component of the corresponding root or eigenvalue, $\lambda_i$. The described periodic behavior moves along a linear manifold defined by a deviation vector from the equilibrium 
\begin{gather}
\delta \vec{X}_{\lambda_i}=\frac{\textbf{\textit{Re}}(\vec{u}_i)}{||\textbf{\textit{Re}}(\vec{u}_i)||_2}\cos{\theta_i} + \frac{\textbf{\textit{Im}}(\vec{u}_i)}{||\textbf{\textit{Im}}(\vec{u}_i)||_2}\sin{\theta_i}
\end{gather}
Where $\vec{u}_i$ are the eigenvectors corresponding to a given eigenvalue.
\\
For the planar system we have already identified one fundamental frequency in the form of the equilibrium orbit rate constraint, $\dot{\theta}^*$. The eigen decomposition provides an additional three oscillations, or fundamental frequencies. Each is associated with one of the planar states; that is the relative separation $R$, and the primary and secondary phase angles, $\phi_1$ and $\phi_2$. Because the orbit angle $\theta$ is an ignorable coordinate it has a zero eigenvalue and thus no associated frequency from the eigen decomposition. In Table 2 the linear periods for the 617 Patroclus system are listed.

\begin{table}[]
\centering
\caption{Linear periods of manifolds about the planar doubly synchronous equilibrium evaluated for the 617 Patroclus system.}
\begin{tabular}{| c | c |}
\hline
\textbf{Manifold}&\textbf{Linear Period [days]}\\\hline
\textbf{$r$}&3.93\\\hline
\textbf{$\phi_1$}&13.60\\\hline
\textbf{$\phi_2$}&12.12\\\hline
\textbf{Orbit}&4.41\\\hline
\end{tabular}
\end{table}

In the linear form, the center manifolds each form an elliptic oscillation of arbitrary amplitude about their associated state with a unique linear period (Table 2). Of interest in the linear system is the influence of the mass parameters on the linear periods of each manifold. To explore this we evaluate the linear periods of the system as the density, volume and axes of each body are scaled, shown in Fig. 3-5. 

\begin{figure}[!htb]
    \centering
    \includegraphics[width=.75\textwidth]{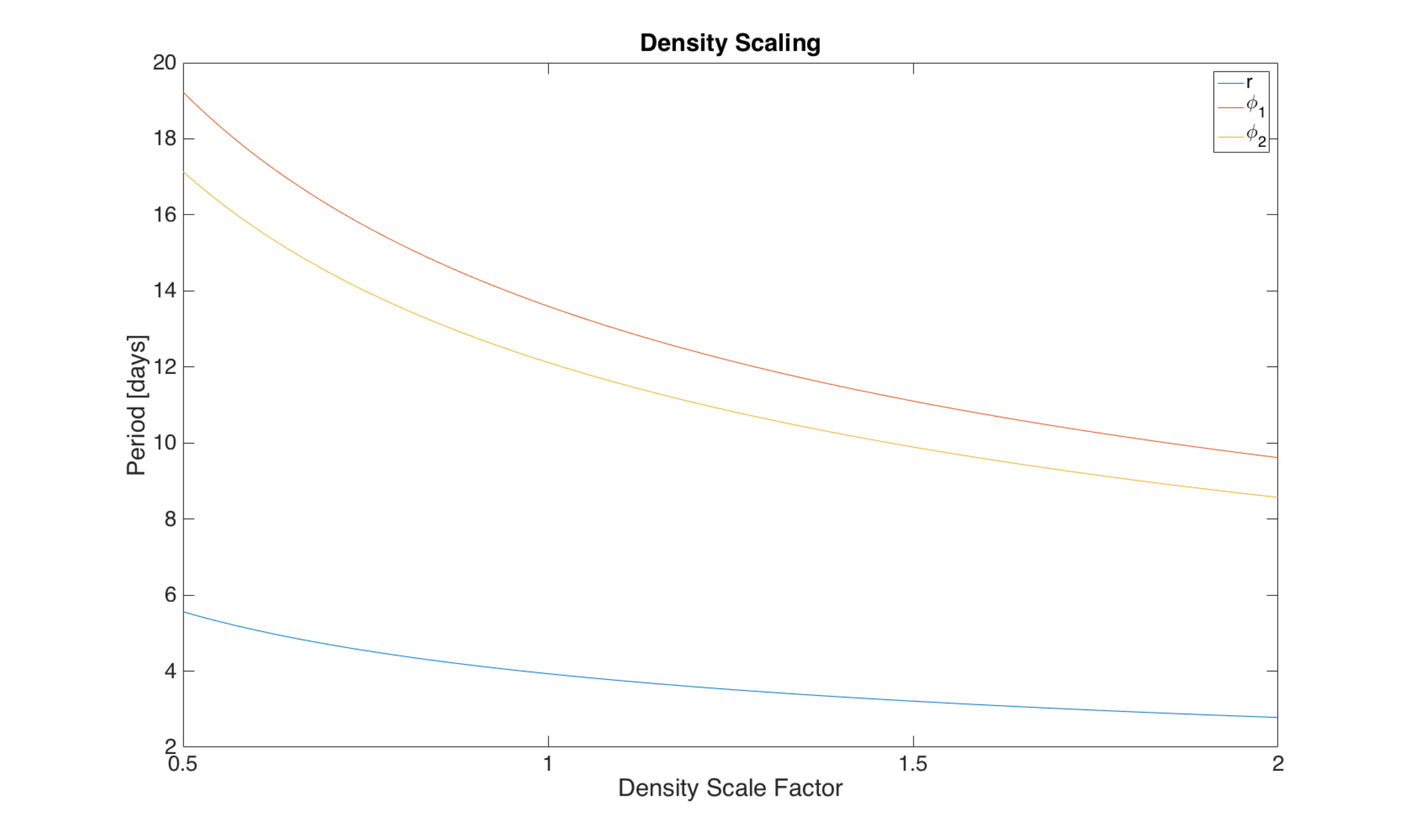}
    \caption{Behavior of planar doubly synchronous manifold linear periods as the density of the system is scaled.}
\end{figure}
\begin{figure}[!htb]
    \centering
    \includegraphics[width=.75\textwidth]{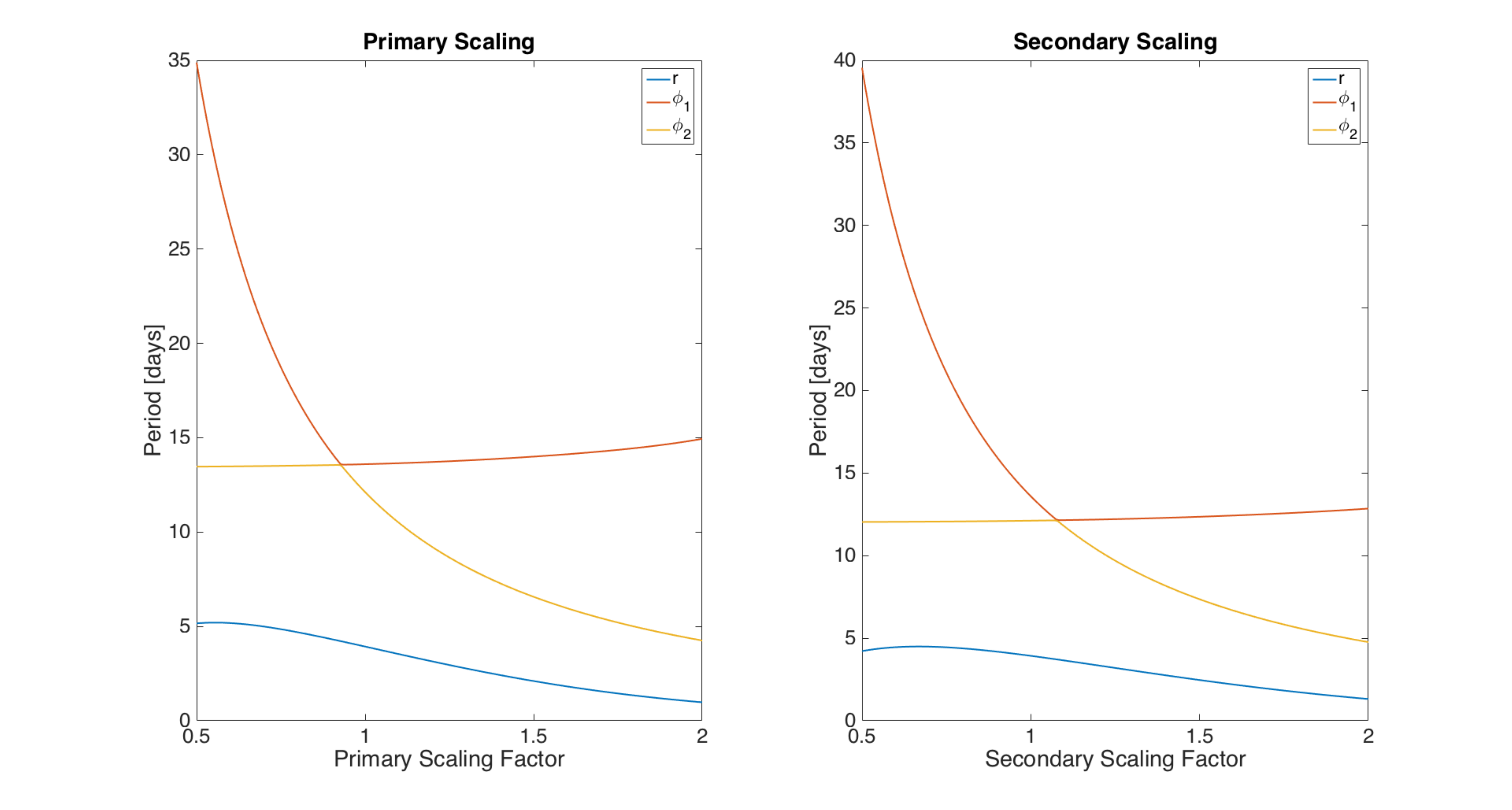}
    \caption{Behavior of planar doubly synchronous manifold linear periods as the volume of each body is scaled.}
\end{figure}
\begin{figure}[!htb]
    \centering
    \includegraphics[width=.75\textwidth, angle=-90]{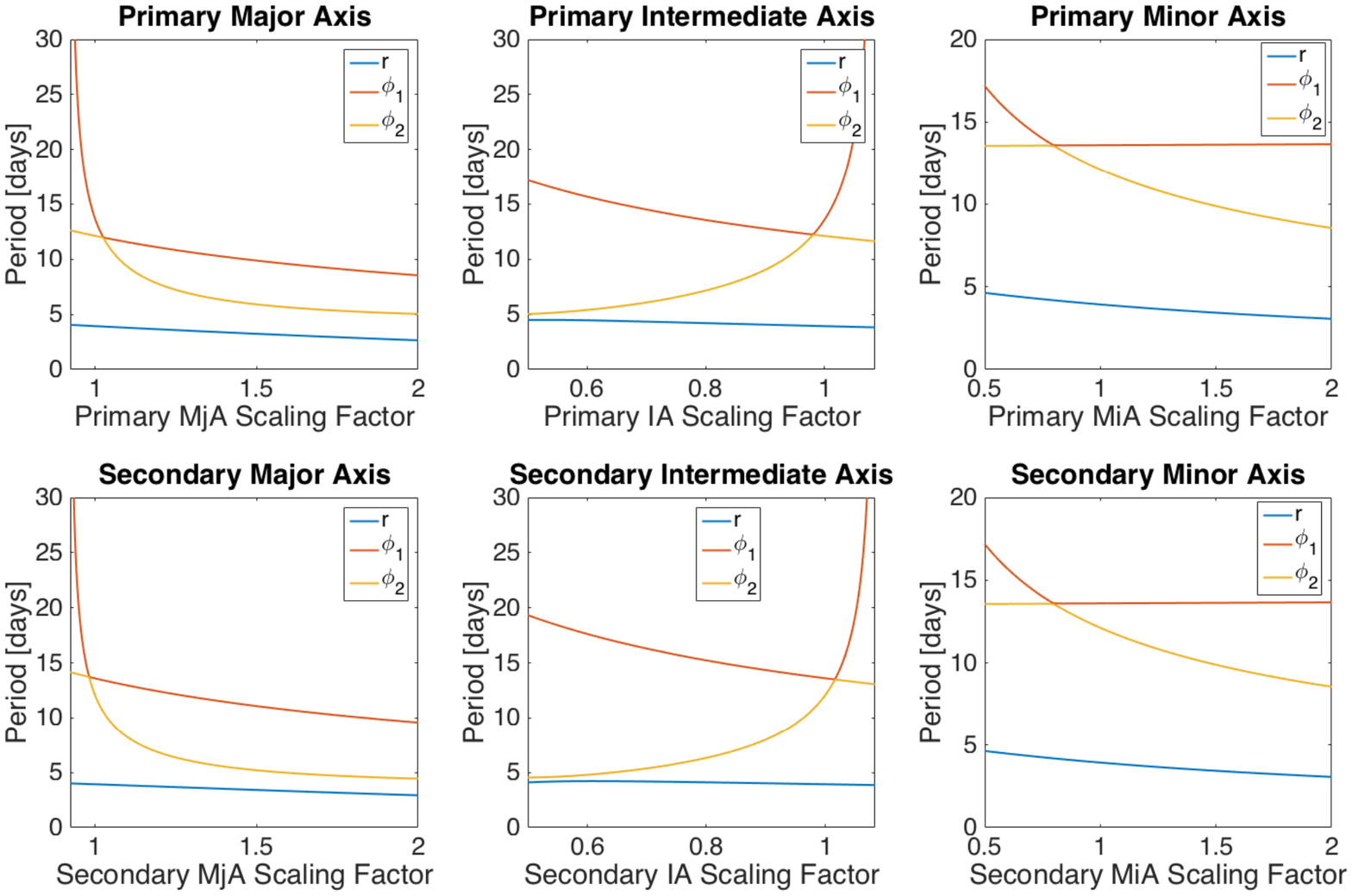}
    \caption{Behavior of planar doubly synchronous manifold linear periods as the length of each ellipsoid axis is scaled.}
\end{figure}

As the density is scaled upwards in Fig. 3 the periods each decrease in length implying a link between the system mass and the speed of its motion. The two phase angle periods also show very similar behavior with a roughly constant gap between their periods, a unique behavior associated with density scaling. In the case of volumetric scaling, Fig. 4, the separation and secondary phase angle periods show a muted response, while the behavior of the primary phase angle period shows a high sensitivity. In this figure, and the axis scaling figure, Fig. 5, there is an apparent switching of the primary and secondary phase angle periods near the unity scaling factor. This results from the secondary becoming more massive as one of the bodies is scaled resulting in the gravitational dominance switching from the primary to the secondary. It is also of note that the scaling of the minor and intermediate axes is truncated near the scaling factor of one, this is done to avoid degeneration of the system by scaling the body to a oblate spheroidal shape. In the axial scaling analysis, Fig. 5, as each of the three axes is independently scaled, regardless of the body, a unique response occurs in the periods for each axis scaling; implying unique behavior associated with each moment of inertia. While this analysis does not provide a definite method of determining the observability of the mass parameters, it does point towards unique behaviors of the system as different aspects of the mass distribution are scaled.  We thus conclude that oscillations about the equilibrium likely will impact observations of binary systems to a significant degree.

\subsection{Planar Observability}
To understand the influence of the mass parameters on the fundamental frequencies, we continue with analysis of the planar and second-order F2BP. The periods of the fundamental frequencies are considered as idealized measurements made by observers. While direct observations of these frequencies would require extensive observations in close proximity, the frequencies more succinctly contain the same information content as direct observations of the dynamics and help to provide a best case analysis. We approximate the sensitivity of the fundamental frequencies to the mass parameters as 

\begin{gather}
\delta \vec{\Omega}=\frac{\partial \vec{\Omega}}{\partial \vec{T}} \bullet \delta \vec{T}
\end{gather}

Where the vector $ \vec{\Omega}$ is the set of fundamental frequencies derived from the eigen decomposition and the doubly synchronous orbit rate
\begin{gather}
\vec{\Omega}=\begin{bmatrix}
 \vec{\Lambda},\dot{\theta}
\end{bmatrix}, \mbox{ where }\vec{\Lambda}=\begin{bmatrix}
\beta_{R},\beta_{\phi_1},\beta_{\phi_2}
\end{bmatrix}
\end{gather}
with $\beta_i$ representing the linearized manifold periods, $P_{\beta_i}$. The vector $\vec{T}$ is the set of second order principal-axis inertia integrals
\begin{gather}
\vec{T}=\begin{bmatrix}
T_A^{2,0,0},T_A^{0,2,0},T_A^{0,0,2},T_B^{2,0,0},T_B^{0,2,0},T_B^{0,0,2}
\end{bmatrix}
\end{gather}
It is assumed that knowledge about the mass of each body is gained from its relative separation and the reflex motion of the system about its center of mass. We define $\frac{\partial \vec{\Omega}}{\partial \vec{T}}$ as the sensitivity matrix, with which a least norm differential corrector can be used to estimate the mass parameters based on the observed frequencies.
\\
The three frequencies arising from the manifold analysis are not analytically derived because of the complexity of the dynamics matrix and instead were computed numerically. Because of this, the partials of these frequencies could not be derived in closed form. Instead we leverage properties of the characteristic equation 
\begin{gather}
|\textbf{\textit{A}}-\lambda_i \textbf{\textit{I}}|=0\\
0=a_n\lambda^n_i +a_{n-1}\lambda^{n-1}_i+...+a_1\lambda_i+a_0
\end{gather}
Where the coefficients $a_n$ represent coefficients found from the determinant computation. We then take the partial of the characteristic equation in its polynomial form with respect each element $j$ of the mass parameter vector, $\vec{T}$
\begin{gather}
0=\frac{\partial a_n}{\partial \vec{T}(j)}\lambda^n_i+a_n\frac{\partial \lambda_i}{\partial \vec{T}(j)}\lambda^{n-1}_i+\frac{\partial a_{n-1}}{\partial \vec{T}(j)}\lambda^{n-1}_i+a_{n-1}\frac{\partial \lambda_i}{\partial \vec{T}(j)}\lambda^{n-2}_i+...\\
\nonumber +\frac{\partial a_1}{\partial \vec{T}(j)}\lambda_i+a_1\frac{\partial \lambda_i}{\partial \vec{T}(j)}
\end{gather}
We can then solve for the the partial $\frac{\partial \lambda_i}{\partial \vec{T}(j)}$ by substituting the numerically generated values of $\lambda$. Using the imaginary component of this partial, we can compute the matrix $\frac{\partial \vec{\beta}}{\partial \vec{T}}$ for use in the sensitivity matrix.

For the planar and second-order realization of the problem the sensitivity matrix is rank deficient at rank 4 while having 6 columns from vector $\vec{T}$. As a result the differential corrector will find a solution lying in a two dimensional solution plane, defined by two nullspace vectors of the sensitivity matrix. Because the mass parameters of interest are the second order principal-axis inertia integrals, we can utilize the definition of the inertia ellipsoid
\begin{gather}
I_{zz} \leq I_{xx} +I_{yy}\\
I_{zz} \geq I_{yy} \geq l_{xx}
\end{gather}
and thus constrain the valid area on the solution plane.

If such an estimation scheme were to be used, then other system observations and measurements would need to be gathered to find an exact solution. For instance flybys of either asteroid measuring the spherical harmonics or other mass tracking techniques. In the case of spherical harmonics measurements, constraints can be derived from the relationship\cite{scheeres2016orbital}

\begin{gather}
I_{x} - I_{y} = - 4 M r_s^2 C_{22}\\
I_{y} - I_{z} = r_s^2 (C_{20}-2C_{22})
\end{gather}

Where $r_s$ is an arbitrary scaling length. The projection of these constraint lines onto the solution plane provides the information necessary to reduce the estimated solution from a two dimensional space to a single point. 

We illustrate the planar estimation approach in Fig. 6 by computing a set of initial mass parameter guesses with a Gaussian perturbation up to 5$\%$ from the ``truth'' mass parameter values. These are then projected onto the solution plane, shown as the red points, where the axes are the two nullspace vectors. We then project the inertia ellipsoid constraints onto the solution plane as the sets of magenta and green lines, for the primary and secondary respectively. Finally, the spherical harmonics constraints are projected onto the solution plane as the blue and black lines for the primary and red and cyan lines for the secondary.Because of the similarity in shape between Patroclus and the secondary Menoetius, the constraint lines appear to be overlapping in the figure. For a more dissimilar binary pair the spherical harmonics constraint lines would be more distinct. With all of these bounds and constraints in the figure, the different steps of this estimation process are clear; resulting in the final green truth at the crossing of the spherical harmonics constraints.
 
\begin{figure}[!htb]
    \centering
    \includegraphics[width=.75\textwidth, angle=-90]{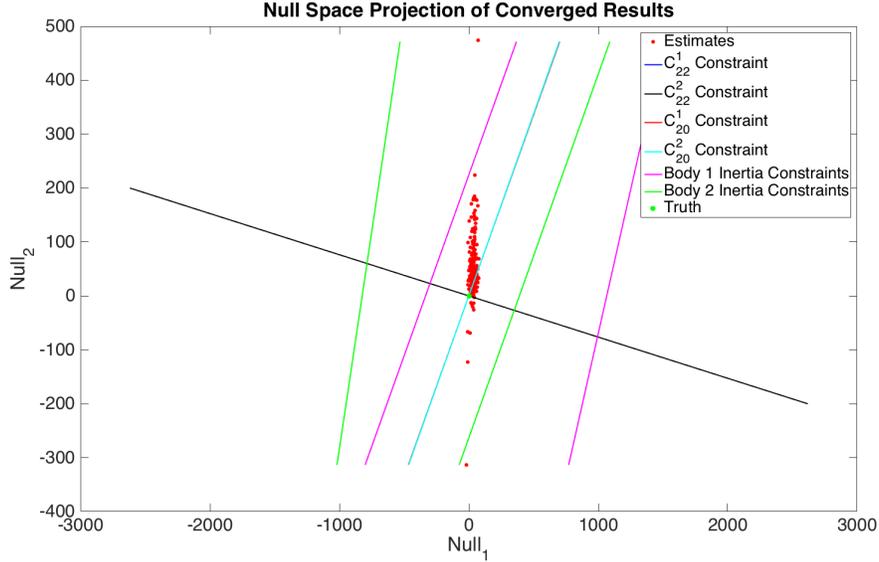}
    \caption{Projection of planar mass parameter estimates onto the solution nullspace plane. The inertia constraints illustrate the bounds caused by the inertia ellipsoid projected onto the plane. The $C_{20}$ and $C_{22}$ constraints illustrate the knowledge added by measurement of either bodies second order spherical harmonics terms; because of the similar shape of the two bodies simulated the spherical harmonics constraints overlap. The axes represent unitless perturbations along each nullspace vector.}
\end{figure}

\section{Nonplanar Mass Parameter Observability}
\subsection{Expansion To Nonplanar Manifolds}
The restrictions of the planar system enable a simple, but limited, approach to detailed analysis of the observability. However understanding the impact of out-of-plane motion on the number of manifolds and their behavior is vital for a full understanding of the mass parameter observability. We thus expand our dynamics model to the more general model with the mutual gravity potential truncated at order two, recalling Eq. 11-13. The increased dimensionality of the nonplanar system results in seven system center manifolds from the eighteen states, as opposed to the three system manifolds from eight states in the planar problem. There are also two ignorable coordinates associated with zero-valued eigenvalues as opposed to the single ignorable coordinate of the planar case. Because of the complexity of developing the mutual gravity potential and torques as well as the equations of motion we choose not to reduce the state representation to a minimal set. Because the doubly synchronous behavior does not change from the planar to nonplanar dynamics the $\dot{\theta}$ constraint remains unchanged. The behaviors of each of these eight fundamental frequencies are illustrated in Fig. 7. 
\begin{figure}[!htb]
    \centering
    \includegraphics[trim={0 5 0 5},width=1.\textwidth, angle=0]{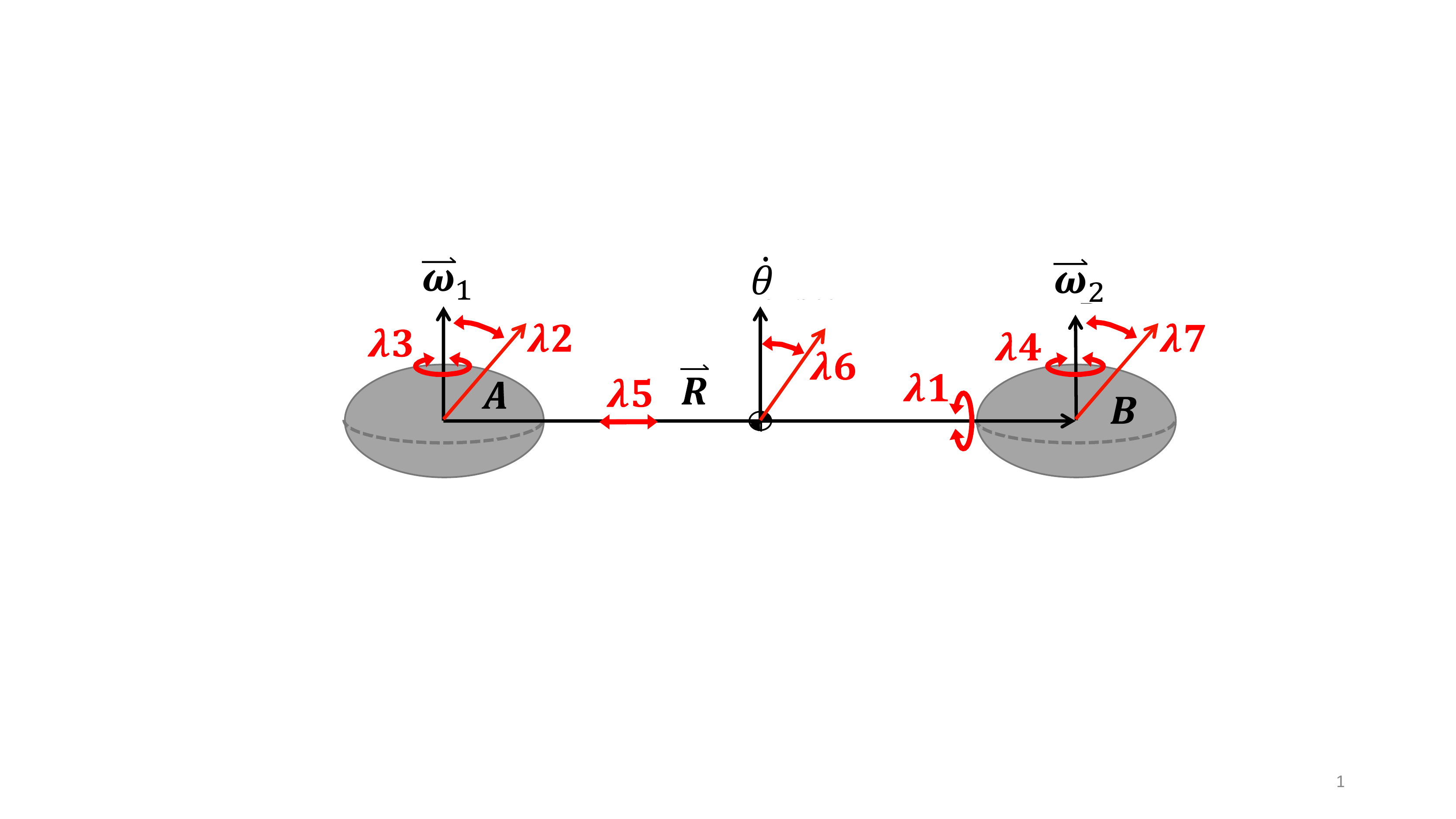}
    \caption{Diagram of the nonplanar doubly synchronous equilibrium. The perturbations associated with each system center manifold are illustrated to visualize their behavior.}
\end{figure}

We see that the three manifolds from the planar case remain for the nonplanar case, now being referred to as $\beta_3$, $\beta_4$, and $\beta_5$. Two of the new manifolds are associated with precession and nutation of the primary and secondary, respectively $\beta_2$ and $\beta_7$. The remaining two new manifolds are associated with a relative axial twist about the radial axis, $\beta_1$, and the precession and nutation of the orbit plane, $\beta_6$. The periods of the linear manifolds for these fundamental frequencies are shown for the 617 Patroclus system in Table 3.
\begin{table}[]
\centering
\caption{Linear periods of manifolds about the nonplanar doubly synchronous equilibrium evaluated for the 617 Patroclus system.}
\begin{tabular}{| c | c |}
\hline
\textbf{Manifold}&\textbf{Linear Period [days]}\\\hline
\textbf{$\beta_1$}&18.97\\\hline
\textbf{$\beta_2$}&15.67\\\hline
\textbf{$\beta_3$}&13.65\\\hline
\textbf{$\beta_4$}&12.13\\\hline
\textbf{$\beta_5$}&4.41\\\hline
\textbf{$\beta_6$}&3.86\\\hline
\textbf{$\beta_7$}&2.91\\\hline
\textbf{Orbit}&4.41\\\hline
\end{tabular}
\end{table}

As with the planar case, for observations of the nonplanar system, we make the assumption that the reflex motion has been well characterized, providing constraints on the relative separation and the mass ratio, defined here as
\begin{gather}
\mu = \frac{M_B}{M_A+M_B}
\end{gather}
Thus the mass of each body can be evaluated based on an estimate of the total mass
\begin{gather}
M_T = M_A+M_B
\end{gather}
As a result our estimated mass parameters can be the total mass and the second order inertia integrals for both bodies.
\begin{gather}
\vec{T}=\begin{bmatrix}
M_T,&T_A^{2,0,0},T_A^{0,2,0},T_A^{0,0,2},T_B^{2,0,0},T_B^{0,2,0},T_B^{0,0,2}
\end{bmatrix}
\end{gather}
Given the increase in the number of system manifolds the system frequency vector, $\vec{\Omega}$, for the nonplanar estimation becomes
\begin{gather}
\vec{\Omega}=\begin{bmatrix}
\beta_{1},\beta_{2},\beta_{3},\beta_{4},\beta_{5},\beta_{6},\beta_{7},\dot{\theta}
\end{bmatrix}
\end{gather}
For this estimation we now have more observables than estimated values, thus the problem is overconstrained and can provide a full solution for the mass parameters from a theoretical standpoint. 

The process of estimation will again apply Eq. 43, however the larger and more complex dynamics matrix for the nonplanar problem requires a new approach to computing the partials of the fundamental frequencies with respect to the mass parameters. Properties of the left and right eigenvectors are leveraged to compute the frequency partials. The left eigenvectors are defined as 
\begin{gather}
\lambda_i \vec{v}_i = \textbf{\textit{A}}^T(\vec{T})\vec{v}_i
\end{gather}
while the right eigenvectors are defined as
\begin{gather}
\lambda_i \vec{u}_i = \textbf{\textit{A}}(\vec{T})\vec{u}_i
\end{gather}
where the $i$ indicates the specific eigenvalue-vector pair. To begin the derivation, the partial of the right eigenvalue equation  is taken with respect to the $j$th mass parameter
\begin{gather}
\frac{\partial \lambda_i}{\partial \vec{T}(j)}\vec{u}_i+\lambda_i\frac{\partial \vec{u}_i}{\partial \vec{T}(j)}=\frac{\partial \textbf{\textit{A}}}{\partial \vec{T}(j)}\vec{u}_i+\textbf{\textit{A}}\frac{\partial \vec{u}_i}{\partial \vec{T}(j)}
\end{gather}
Left multiplying this partial by the transpose of the left eigenvectors, $\vec{v}_i^T$, the equation becomes
\begin{gather}
\frac{\partial \lambda_i}{\partial \vec{T}(j)}\vec{v}_i^T\vec{u}_i+\lambda_i\vec{v}_i^T\frac{\partial \vec{u}_i}{\partial \vec{T}(j)}=\vec{v}_i^T\frac{\partial \textbf{\textit{A}}}{\partial \vec{T}(j)}\vec{u}_i+\vec{v}_i^T\textbf{\textit{A}}\frac{\partial \vec{u}_i}{\partial \vec{T}(j)}
\end{gather}
in which the $\frac{\partial \vec{u}_i}{\partial \vec{T}(j)}$ terms cancel based on the definition of the left eigenvector. Rearranging to solve for the partial of the frequency, $\beta_i$ only
\begin{gather}
\frac{\partial \beta_i}{\partial \vec{T}(j)}=\textbf{\textit{Im}}\Big(\frac{1}{\vec{v}_i^T\vec{u}_i} \vec{v}_i^T\frac{\partial \textbf{\textit{A}}}{\partial \vec{T}(j)}\vec{u}_i\Big)
\end{gather}
Thus the sensitivity matrix for the nonplanar problem can be computed element by element, iterating over this equation.

Given the complete solution generated by the estimation process, we can now analyze the uncertainties of the estimated mass parameters. The covariance and correlation of the mass parameters generated from this approach help to quantify the influence of the mass parameters on dynamical observations. To begin this analysis the covariance matrix for the mass parameters, $P_{TT}$, is formulated by relating the pseudo-inverse of the sensitivity matrix and the observational covariance of the fundamental frequencies, $\textbf{\textit{P}}_{\Omega\Omega}$
\begin{gather}
\textbf{\textit{P}}_{TT}=\delta \vec{T} \delta \vec{T}^T=\Big(\frac{\partial \vec{\Omega}}{\partial \vec{T}} ^T\frac{\partial \vec{\Omega}}{\partial \vec{T}}\Big)^{-1}\frac{\partial \vec{\Omega}}{\partial \vec{T}}^T \bullet \textbf{\textit{P}}_{\Omega\Omega} \bullet \frac{\partial \vec{\Omega}}{\partial \vec{T}}\Big(\frac{\partial \vec{\Omega}}{\partial \vec{T}} ^T\frac{\partial \vec{\Omega}}{\partial \vec{T}}\Big)^{-1}
\end{gather}
Because the frequencies are not an intuitive measurement the covariance of the fundamental frequencies is converted to the covariance of fundamental periods
\begin{gather}
\textbf{\textit{P}}_{\Omega\Omega}=\frac{\partial \vec{\Omega}}{\partial \vec{P}}\bullet \textbf{\textit{P}}_{PP}\bullet \frac{\partial \vec{\Omega}}{\partial \vec{P}}^T
\end{gather}
This conversion is simply the derivative of the frequency and period relationship.
\begin{gather}
\frac{\partial \vec{\Omega}}{\partial \vec{P}}=\frac{\textbf{\textit{diag}}\big(\vec{\Omega}\big)^2}{2\pi}
\end{gather}
To define the covariance of the fundamental periods it is assumed that all periods would be independently measured using the same observation technique.
\begin{gather}
\textbf{\textit{P}}_{PP}=\sigma_P^2 \bullet \textbf{\textit{I}}
\end{gather}
where there is a single period observational variance, $\sigma_P$, that is applied to the observations of each period. This observational variance can be considered to be a temporal resolution of the period measurements or the precision of each measurement.
\subsection{An Idealized Estimation Method}
This formulation of the covariance allows us to treat the sensitivity analysis approach as an idealized estimation approach. Here the observational variance, $\sigma_P$, would act as an observational accuracy requirement to constrain the mass parameter estimates to a specific accuracy level. In Table 4 this is leveraged to predict the observational variance necessary in order to gain 10$\%$ knowledge of each of the seven mass parameters of the 617 Patroclus system. 
\begin{table}[]
\centering
\caption{Allowable observational variance of observations for 10$\%$ certainty of mass parameter estimates of the 617 Patroclus system.}
\begin{tabular}{| c | c |}
\hline
\textbf{Mass Parameter}&\textbf{Observational Accuracy Requirement [sec]}\\\hline
\textbf{$M_T$}&10584.0\\\hline
\textbf{$T^{200}_A$}&25.3\\\hline
\textbf{$T^{020}_A$}&25.3\\\hline
\textbf{$T^{002}_A$}&25.3\\\hline
\textbf{$T^{200}_B$}&15.0\\\hline
\textbf{$T^{020}_B$}&15.0\\\hline
\textbf{$T^{002}_B$}&15.0\\\hline
\end{tabular}
\end{table}

To provide further insight, we compute the mass parameter variance for 617 Patroclus, given a 1 second observation variance 
\begin{gather}
\bar{\sigma}_{T}=\left[\begin{matrix}
9.45\times 10^{-6}&  3.95\times 10^{-3}&  3.94\times 10^{-3}&  3.95\times 10^{-3} \end{matrix}\right.\\
\nonumber \qquad\qquad\qquad\qquad\left.\begin{matrix}6.64\times 10^{-3}&  6.64\times 10^{-3}&  6.64\times 10^{-3}
\end{matrix} \right]^T
\end{gather}
and the correlation matrix 
\begin{gather}
\textbf{\textit{$\rho$}}_{TT}=\begin{bmatrix}
  1.& -0.27&-0.27&-0.27&-0.64&-0.64&-0.64\\
       -0.27&  1.&0.99&  0.99& -0.55& -0.55 & -0.55 \\
        -0.27& 0.99&1.& 0.99&  -0.55&  -0.55&-0.55\\
        -0.27&0.99& 0.99&  1.& -0.55&  -0.55&-0.55\\
        -0.64& -0.55&-0.55& -0.55&  1.& 0.99&0.99\\
        -0.64& -0.55 &-0.55&-0.55&0.99&  1.& 0.99  \\
        -0.64& -0.55 &-0.55&-0.55&0.99&  0.99&  1.
\end{bmatrix}
\end{gather}
The values for the variance and correlation in Eq. 66 and 67 are generated from the covariance matrix, which is not included for the sake of brevity and clarity. The bar used with variance values are normalized by the mass parameters corresponding to each element, such that they represent fractional covariances and variances
\begin{gather}
\bar{\textbf{\textit{P}}}_{TT}(i,j)=\frac{P_{TT}(i,j)}{\vec{T}(i)\vec{T}(j)}\\
\bar{\sigma}_{T}(i)=\frac{\sigma_{T}(i)}{\vec{T}(i)}\\
\textbf{\textit{$\rho$}}_{TT}(i,j)=\frac{\bar{\textbf{\textit{P}}}_{TT}(i,j)}{\bar{\sigma}_{T}(i) \bar{\sigma}_{T}(j)}
\end{gather}

\subsection{Mass Parameter Scaling Effects on Observability}
While these results show the mass parameters to be well estimated based on their variances, the observational requirement of 1 second variance on each frequency is highly restrictive. Likewise, the observational requirements to achieve 10$\%$ knowledge of the inertia integrals in table 4 further illustrates the accuracy of observations necessary for second order parameter measurements. In combination with the high correlation on the second order parameters, it becomes clear that without further constraints, via in-situ gravity measurement or other methods, such an estimation approach is not feasible even with these idealized observations. 

To better understand the observability of binary systems as a whole we now investigate the effects of scaling the mass parameters on the correlation and covariance matrices. Specifically we scale the mass ratio and second order principal inertia integrals of Menoetius, the 617 Patroclus secondary. This is done as a means to explore the effects of mass ratio and differing asteroid mass distributions on the mass parameter observability. We do not simultaneously scale the second order asteroid inertia integrals as this would merely scale the correlation and covariance as opposed to changing their structure. Likewise only the secondary is scaled as the system is nearly symmetric such that scaling either body will show the same observability behavior. The scaling of the inertia integrals is limited such that the asteroid remains a triaxial ellipsoid as opposed to an oblate spheroid; this is to avoid the degeneration of the manifolds of the system caused by symmetric semi-axes. 

For simplicity the second order principal axis inertia integrals of the secondary are refered to here as $T_x$, $T_y$, and $T_z$. The study performed scales the mass ratio by .1 and .55, $T_x$ by .88 and .94, $T_y$ by .7 and .85, and $T_z$ by .01 and .51.  The scaling floor for $T_x$ and $T_y$ is selected to avoid the the inertia integral dropping below the next smallest inertia integral which would cause an spherical degeneracy, thus $T_z$ has no floor. Fig. 8 and 9 illustrate the covariance and correlation with the covariance elements colored by their log value and the correlation elements colored by their linear value from -1 to 1. The key point of interest in this case is whether the scaling of these mass parameters can lower the correlation between the each body's second order inertia integrals. What both results show is that outside of the case of an extremely flattened body, the second order inertia integrals remain highly correlated. Even in the case of an extremely flattened body only the $T_z$ correlation changes significantly while the $T_x$ and $T_y$ relationship remains very coupled. This implies that the dynamical effects of these unknown mass parameters can be significant for most shapes and configurations.

\begin{figure}[!htb]
    \centering
    \includegraphics[width=.75\textwidth, angle=0]{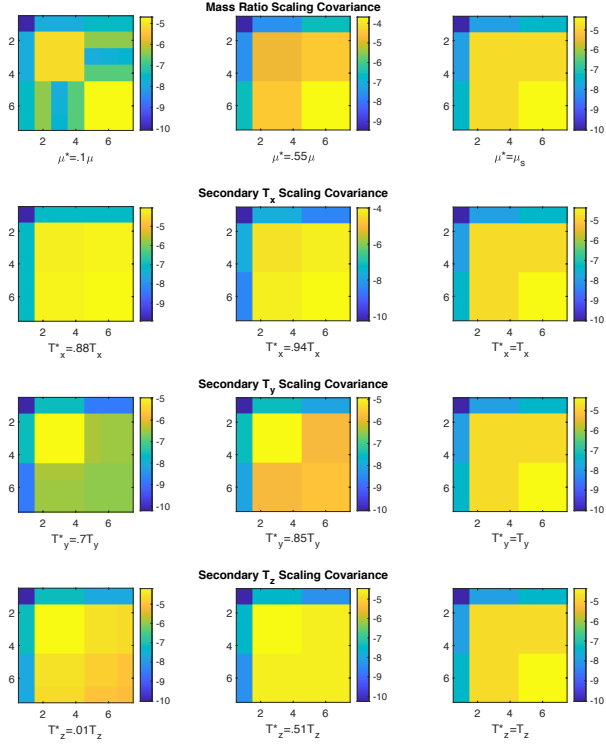}
    \caption{Effects of secondary mass parameter scaling on mass parameter covariance matrix with magnitude of matrix elements shown in log color.}
\end{figure}

\begin{figure}[!htb]
    \centering
    \includegraphics[width=.75\textwidth, angle=0]{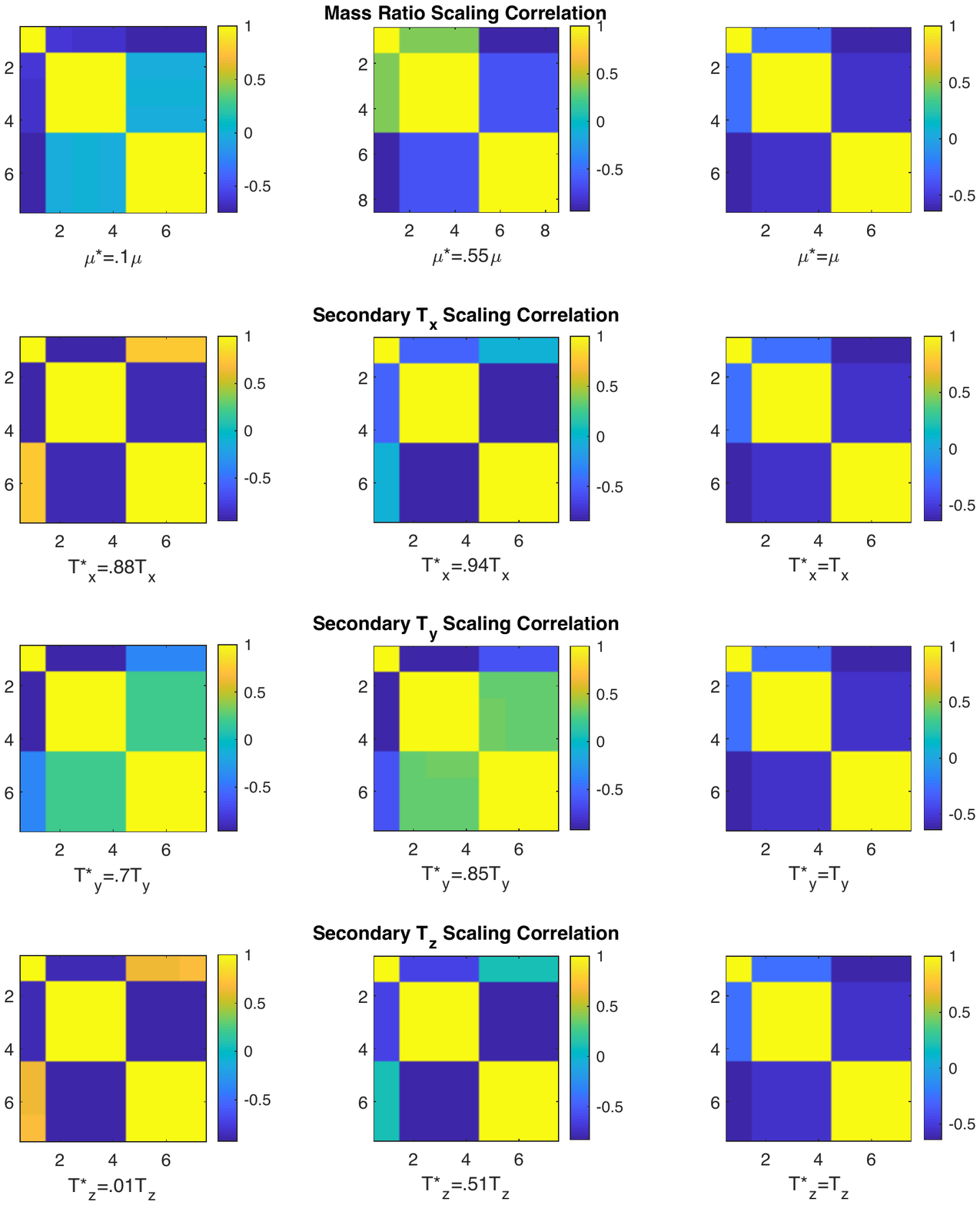}
    \caption{Effects of secondary mass parameter scaling on mass parameter correlation matrix with magnitude of matrix elements shown in linear color.}
\end{figure}

\section{Near-Spherical Doubly Synchronous Systems}
For a number of observed doubly synchronous binaries current shape knowledge is limited to a mean radius value or spherical shape estimate. However, for a binary system to remain in a stable doubly synchronous orbit, dynamical analysis has shown that the mass distribution of the bodies must be elongated such that the bodies mutual gravity torques exist to enforce the tidal locking. One result of this is that for a sphere-sphere doubly synchronous system four of the manifolds become zero eigenvalues due to the lack of attitude interaction between the bodies, described in Table 5 for the Pluto-Charon system. In addition, for a nearly spherical body the effects of any elongation will be so low that the periods of these four manifolds will be functionally immeasurable.

To better analyze these near spherical systems we perform a linearization about the spherical shape of the bodies using an elongation factor $\epsilon$ which perturbs the bodies as triaxial ellipsoids defined such that
\begin{gather}
abc=R^3\\
a=R(1+\epsilon), \quad b=R, \quad c=R(1-\epsilon)\\
I_x=\frac{2MR^2}{5}(1-\epsilon), \quad I_y=\frac{2MR^2}{5}, \quad I_x=\frac{2MR^2}{5}(1+\epsilon)
\end{gather}
where $a$, $b$, and $c$ are the semi-axes. 

Using these definitions of the near-spherical mass distributions we can linearize the dynamics as 
\begin{gather}
\delta \vec{\Omega} =\Big[ \frac{\partial \vec{\Omega}}{\partial \vec{\epsilon}}\Big]_{sphere} \delta \vec{\epsilon}
\end{gather}
\begin{gather}
\delta \vec{\Omega} =\Big[ \frac{\partial \vec{\Omega}}{\partial \vec{T}} \frac{\partial \vec{T}}{\partial \vec{\epsilon}}\Big]_{sphere} \delta \vec{\epsilon}
\end{gather}
\begin{gather}
\frac{\partial \vec{T}}{\partial \vec{\epsilon}} =\begin{bmatrix}
\frac{-2R_P^2}{5} & 0\\
0&0\\
0&\frac{2R_P^2}{5}\\
\frac{-2R_C^2}{5} & 0\\
0&0\\
0&\frac{2R_C^2}{5}
\end{bmatrix}
\end{gather}
where $\vec{\Omega}$ represents the remaining system frequenciess, $\vec{T}$ again represents the vector of only the second-order principal-axis inertia integrals, and $\vec{\epsilon}$ represents the elongation factors applied to the primary and secondary body independently. It is of note that this linearization assumes that the equilibrium separation remains constant as the bodies are elongated. This means that the equilibrium orbit rate must scale as the bodies are elongated while maintaining the observed separation. This effect on the orbit rate is analogous to the effect on the other three periods of the elongation, however they are not as simply expressed as the orbit rate, whose change can be directly computed. By analyzing the linearized effects of the deformation on the four remaining periods and observing the behavior of these measurable manifold periods, we provide a different approach to understand the effects of the mass distributions of near-spherical systems.

\subsection{Application to Pluto-Charon System}
To illustrate the use of this approach we apply it to the Pluto-Charon system, a doubly synchronous binary for which only mean radius information has been reliably measured. The density and shape results of Nimmo et al.'s analysis of New Horizons images report the density of Pluto and Charon to be 1854 $\frac{kg}{m^3}$ and 1701 $\frac{kg}{m^3}$ and the mean radii to be 1188.3$\pm$1.6 km and 606.0$\pm$1.0 km respectively\cite{nimmo2017mean}. Applying our analysis to these parameters the system periods can be computed for the spherical system, Table 5.

\begin{table}[]
\centering
\caption{Linear periods of manifolds about the nonplanar doubly synchronous equilibrium evaluated for the spherical Pluto-Charon system.}
\begin{tabular}{| c | c |}
\hline
\textbf{Manifold}&\textbf{Linear Period [days]}\\\hline
\textbf{$\beta_1$}&DNE\\\hline
\textbf{$\beta_2$}&DNE\\\hline
\textbf{$\beta_3$}&DNE\\\hline
\textbf{$\beta_4$}&DNE\\\hline
\textbf{$\beta_5$}&6.39\\\hline
\textbf{$\beta_6$}&6.39\\\hline
\textbf{$\beta_7$}&6.39\\\hline
\textbf{Orbit}&6.39\\\hline
\end{tabular}
\end{table}

Beginning from the spherical system periods we apply the linearization in three ways to understand what information can be gained from this analysis. The first approach is to perturb only the shape of Pluto, next only the shape of Charon is perturbed, and finally the shape of both bodies are identically perturbed. This is to say that the vector $\vec{\epsilon}$ can be expressed as $[1,0]\cdot\epsilon_{Pluto}$, $[0,1]\cdot\epsilon_{Charon}$, and $[1,1]\cdot\epsilon_{system}$ respectively. From these three approaches, illustrated in Fig. 10, we see that the relative behavior of the periods differs uniquely for deformation of each body. This implies that through measurement of the relative lengths of the periods, information  on the mass distribution could be gathered to help constrain each body's mass distribution. As is clear from these results however, the deformation of the bodies would need to be sufficiently large to be detected by realistic measurement methods.

\begin{figure}[!htb]
    \centering
    \includegraphics[width=.75\textwidth, angle=-90]{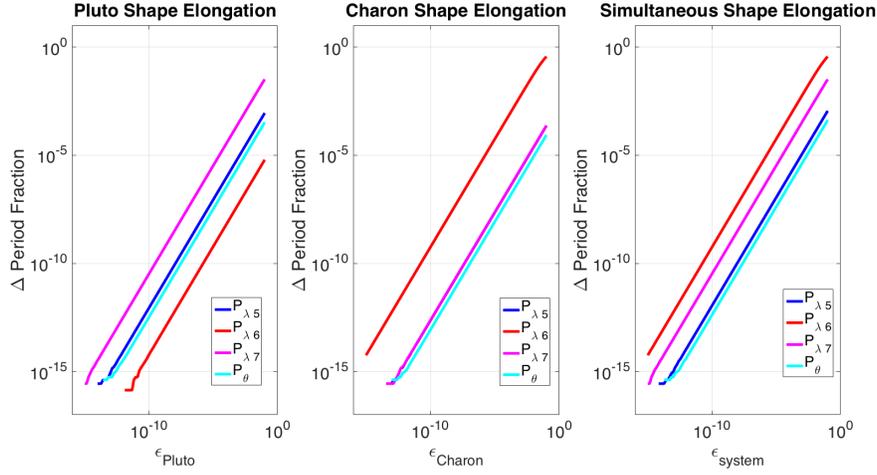}
    \caption{Linearized change in Pluto-Charon manifold periods as shape is modified by $\epsilon$ parameter. From left to right the figures show and elongation of only Pluto, elongation of only Charon, and equal elongation of both Pluto and Charon }
\end{figure}

For a comparison point we also compute the complete set of nonplanar periods associated with an ellipsoidal mass distribution for both Pluto and Charon generated with a value of $\epsilon$=.0008. This value of $\epsilon$ is selected because it lies in the middle of the certainty bounds on the mean radius values reported by Nimmo et al; representing roughly a 1 km deformation in Pluto's semi-axes and a .5 km change in Charon's semi-axes. This is reported in Table 6 and further confirms the difficulty of applying our approach to a system so near to the spherical case. The four short periods would not be feasible to distinguish or measure with sufficient accuracy. The four long periods on the other hand would require logistically impossible measurement efforts to be accurately and precisely observed, due to their length.

 \begin{table}[]
\centering
\caption{Linear periods of manifolds about the nonplanar doubly synchronous equilibrium evaluated for the $\epsilon$=.0008 Pluto-Charon system.}
\begin{tabular}{| c | c |}
\hline
\textbf{Manifold}&\textbf{Linear Period [days]}\\\hline
\textbf{$\beta_1$}&4659.25\\\hline
\textbf{$\beta_2$}&1442.06\\\hline
\textbf{$\beta_3$}&393.35\\\hline
\textbf{$\beta_4$}&96.37\\\hline
\textbf{$\beta_5$}&6.39\\\hline
\textbf{$\beta_6$}&6.39\\\hline
\textbf{$\beta_7$}&6.33\\\hline
\textbf{Orbit}&6.39\\\hline
\end{tabular}
\end{table}

\section{Conclusions and Future Work}
In this paper we have shown that the influence of mass parameters on the observable dynamics of binary asteroids, up to the second order, in the doubly synchronous equilibrium will affect dynamical observations significantly. The use of the doubly synchronous assumption allowed for a relatively simple differential corrector method to elucidate fundamental frequencies of the system as a target binary oscillates near the equilibrium. In order to accomplish this we analyzed the manifolds of the planar problem and investigated the influence of mass parameters on the linear behavior of these manifolds to understand how they may affect observations of these systems. For the planar case we found that the mass parameters were not fully observable based purely on observations of the dynamics and would require other in-situ or remote observations to constrain a system's mass parameters. For the nonplanar F2BP we were able to show that the mass parameters were fully observable using only observations of the system dynamics, although the observational requirements are demanding. For the nonplanar differential corrector estimation we were able to investigate the achievable covariance of the estimated mass parameters based on the accuracy of observations of the system dynamics. This provides an understanding of the information quality requirements for the proposed mass parameter estimation approach to be effective. From this analysis we can conclude that more robust measurements, likely from an in-situ spacecraft, would be necessary for mass parameter estimation. Finally, a limited approach to the application of this analysis to near-spherical systems was presented and applied to the Pluto-Charon system.

\appendix
\section{The Hou Mutual Gravity Potential}
The general order Hou et al. reformulation of the mutual gravity potential begins from the double integral description of the mutual gravity potential.
\begin{gather}
U=-G\sum_{n=0}^{N}\frac{1}{R^{n+1}}\tilde{U}_n
\end{gather}
A binomial expansion and Legendre polynomial expansion are then performed to approximate the mass distribution to order $N$ with the inertia integrals in a recursive summation.
\begin{gather}
\tilde{U}_n=\sum_{k(2)=n}^0 t^n_k\sum_{(i_1,i_2,i_3)(i_41,i_5,i_6)(j_1,j_2,j_3)(j_4,j_5,j_6)}a^k_{(i_1,i_2,i_3)(i_4,i_5,i_6)}\\
\nonumber \times b^{n-k}_{(j_1,j_2,j_3)(j_4,j_5,j_6)} e_x^{i_1+i_4}e_y^{i_2+i_5}e_z^{i_3+i_6}  M_A T_A^{(i_1+j_1),(i_2+j_2),(i_3+j_3)}\\
\nonumber \times M_B T_B^{\prime (i_4+j_4),(i_5+j_5),(i_6+j_6)}\\
\nonumber \text{Where k(2) implies stepping by 2 as opposed to 1}
\end{gather}
Here the recursive coefficient is represented by $t^n_k$.
\begin{gather}
t_{k+2}^n=-\dfrac{(n-k)(n+k+1)}{(k+2)(k+1)}
\end{gather}
While the binomial expansion coefficients are $a^k_{(i_1,i_2,i_3)(i_4,i_5,i_6)}$ and $b^{n-k}_{(j_1,j_2,j_3)(j_4,j_5,j_6)}$.
\begin{gather}
a^k_{(i_1,i_2,i_3)(i_4,i_5,i_6)}=a^{k-1}_{(i_1-1,i_2,i_3)(i_4,i_5,i_6)}+a^{k-1}_{(i_1,i_2-1,i_3)(i_4,i_5,i_6)}\\ \nonumber+a^{k-1}_{(i_1,i_2,i_3-1)(i_4,i_5,i_6)}-a^{k-1}_{(i_1,i_2,i_3)(i_4-1,i_5,i_6)}-a^{k-1}_{(i_1,i_2,i_3)(i_4,i_5-1,i_6)}\\ \nonumber-a^{k-1}_{(i_1,i_2,i_3)(i_4,i_5,i_6-1)}
\end{gather}
\begin{gather}
b^k_{(j_1,j_2,j_3)(j_4,j_5,j_6)}=b^{k-2}_{(j_1-2,j_2,j_3)(j_4,j_5,j_6)}+b^{k-2}_{(j_1,j_2-2,j_3)(j_4,j_5,j_6)}\\ \nonumber+b^{k-2}_{(j_1,j_2,j_3-2)(j_4,j_5,j_6)}+b^{k-2}_{(j_1,j_2,j_3)(j_4-2,j_5,j_6)}+b^{k-2}_{(j_1,j_2,j_3)(j_4,j_5-2,j_6)}+b^{k-2}_{(j_1,j_2,j_3)(j_4,j_5,j_6-2)}\\ \nonumber-2b^{k-2}_{(j_1-1,j_2,j_3)(j_4-1,j_5,j_6)}-2b^{k-2}_{(j_1,j_2-1,j_3)(j_4,j_5-1,j_6)}-2b^{k-2}_{(j_1,j_2,j_3-1)(j_4,j_5,j_6-1)}
\end{gather}
For these coefficients the superscripts and subscripts serve as indices. The $i$ and $j$ expansion indices are constrained by  and summation over these indices sums over the possible combinations of these indices based on the values of $k$ and $n$ and the constraint equations.
\begin{gather}
k=i_1+i_2+i_3+i_4+i_5+i_6\\
n-k=j_1+j_2+j_3+j_4+j_5+j_6
\end{gather}
The variables $R$ and $e_x$, $e_y$, and $e_z$ represent the magnitude and unit direction of the relative separation. The variables $T_A$ and $T^{\prime}_B$ are the mass-normalized inertia integral sets for the primary and secondary bodies, where the prime denotes that the inertia integrals of the secondary are rotated into the frame of the primary. 

\subsection{Inertia Integrals}
Central to this reformulation of the mutual potential is the use of inertia integrals to describe the mass distribution. This aspect of the reformulation is accomplished by the application of a Legendre polynomial expansion to describe the mass distributions, where the Legendre coefficients are referred to as inertia integrals. 
\begin{gather}
T^{l,m,n}=\frac{1}{MR^{l+m+n}}\int_B x^l y^m z^n dm \textit{, where }l+m+n=N
\end{gather}
In this way the inertia integrals can be considered analogous in use to spherical harmonics\cite{tricarico2008figure}. The mathematical form of the inertia integrals is similar to that of the moments and products of inertia for a rigid body wherein each term represents the mass distribution about some axis, however the inertia integrals are expanded to order $N$ whereas moments of inertia are linear combinations of second order inertia integrals. Here we provide the mass and length normalized form of an inertia integral along with the 0th and 2nd order coefficients in terms of the normalized moments and product of inertia.
\begin{gather}
T^{l,m,n}=\frac{1}{MR^{l+m+n}}\int_B x^l y^m z^n dm \textit{, where }l+m+n=N\\
1=T^{0,0,0} \\
I_{xx}=T^{0,2,0} +T^{0,0,2}\\
I_{yy}=T^{2,0,0} +T^{0,0,2}\\
I_{zz}=T^{2,0,0} +T^{0,2,0}\\
I_{xy}=-T^{1,1,0}\\
I_{xz}=-T^{1,0,1}\\
I_{yz}=-T^{0,1,1}
\end{gather}
It is of note that the 0th order inertia integral is equal to the mass for the non-normalized form, thus it is equal to one in the normalized form.

\section{Planar Dynamics Matrix}
\begin{gather}
\vec{X}=\begin{bmatrix}
r&\theta&\phi_1&\phi_2&\dot{r}&\dot{\theta}&\dot{\phi_1}&\dot{\phi_2}
\end{bmatrix}^T
\end{gather}
\begin{gather}
\dot{\vec{X}}=\textbf{\textit{A}}\vec{X}=\begin{bmatrix}
0&0&0&0&1&0&0&0\\
0&0&0&0&0&1&0&0\\
0&0&0&0&0&0&1&0\\
0&0&0&0&0&0&0&1\\
\frac{\partial \ddot{r}}{\partial r} & 0 & \frac{\partial \ddot{r}}{\partial \phi_1} & \frac{\partial \ddot{r}}{\partial \phi_2} & 0 & \frac{\partial \ddot{r}}{\partial \dot{\theta}} & 0 &0\\
\frac{\partial \ddot{\theta}}{\partial r} & 0 & \frac{\partial \ddot{\theta}}{\partial \phi_1} & \frac{\partial \ddot{\theta}}{\partial \phi_2} & \frac{\partial \ddot{\theta}}{\partial \dot{r}} & \frac{\partial \ddot{\theta}}{\partial \dot{\theta}} & 0 &0\\
\frac{\partial \ddot{\phi_1}}{\partial r} & 0 & \frac{\partial \ddot{\phi_1}}{\partial \phi_1} & \frac{\partial \ddot{\phi_1}}{\partial \phi_2} & \frac{\partial \ddot{\phi_1}}{\partial \dot{r}} & \frac{\partial \ddot{\phi_1}}{\partial \dot{\theta}} & 0 &0\\
\frac{\partial \ddot{\phi_2}}{\partial r} & 0 & \frac{\partial \ddot{\phi_2}}{\partial \phi_1} & \frac{\partial \ddot{\phi_2}}{\partial \phi_2} & \frac{\partial \ddot{\phi_2}}{\partial \dot{r}} & \frac{\partial \ddot{\phi_2}}{\partial \dot{\theta}} & 0 &0
\end{bmatrix}\begin{bmatrix}
r\\\theta\\\phi_1\\\phi_2\\\dot{r}\\\dot{\theta}\\\dot{\phi_1}\\\dot{\phi_2}
\end{bmatrix}
\end{gather}
\begin{gather}
\frac{\partial \ddot{r}}{\partial r}=\dot{\theta}^2-\frac{V_{rr}}{m}
\end{gather}
\begin{gather}
\frac{\partial \ddot{r}}{\partial \phi_1}=-\frac{V_{r\phi_1}}{m}
\end{gather}
\begin{gather}
\frac{\partial \ddot{r}}{\partial \phi_2}=-\frac{V_{r\phi_2}}{m}
\end{gather}
\begin{gather}
\frac{\partial \ddot{r}}{\partial \dot{\theta}}=2\dot{\theta}r
\end{gather}

\begin{gather}
\frac{\partial \ddot{\theta}}{\partial r}=-2\frac{V_{\phi1}}{mr^3}+\frac{V_{\phi_1}}{mr^2}-2\frac{V_{\phi2}}{mr^3}+\frac{V_{\phi_2}}{mr^2}+2\frac{\dot{r}\dot{\theta}}{r^2}
\end{gather}
\begin{gather}
\frac{\partial \ddot{\theta}}{\partial \phi_1}=\frac{V_{\phi_1\phi_1}}{mr^2}
\end{gather}
\begin{gather}
\frac{\partial \ddot{\theta}}{\partial \phi_2}=\frac{V_{\phi_2\phi_2}}{mr^2}
\end{gather}
\begin{gather}
\frac{\partial \ddot{\theta}}{\partial \dot{r}}=-2\frac{\dot{\theta}}{r}\\
\frac{\partial \ddot{\theta}}{\partial \dot{\theta}}=-2\frac{\dot{r}}{r}
\end{gather}

\begin{gather}
\frac{\partial \ddot{\phi_1}}{\partial r}=2\frac{\big(V_{\phi_1}+V_{\phi_2}\big)}{mr^3}-\frac{\big(V_{r\phi_1}+V_{r\phi_2}\big)}{mr^2}-\frac{V_{r\phi_1}}{M_A I_{A,zz}}-2\frac{\dot{r}\dot{\theta}}{r^2}
\end{gather}
\begin{gather}
\frac{\partial \ddot{\phi_1}}{\partial \phi_1}=-\frac{V_{\phi1\phi1}}{mr^2}-\frac{V_{\phi1\phi1}}{M_A I_{A,zz}}
\end{gather}
\begin{gather}
\frac{\partial \ddot{\phi_1}}{\partial \phi_2}=-\frac{V_{\phi2\phi2}}{mr^2}
\end{gather}
\begin{gather}
\frac{\partial \ddot{\phi_1}}{\partial \dot{r}}=2\frac{\dot{\theta}}{r}
\end{gather}
\begin{gather}
\frac{\partial \ddot{\phi_1}}{\partial \dot{\theta}}=2\frac{\dot{r}}{r}
\end{gather}

\begin{gather}
\frac{\partial \ddot{\phi_2}}{\partial r}=2\frac{\big(V_{\phi_1}+V_{\phi_2}\big)}{mr^3}-\frac{\big(V_{r\phi_1}+V_{r\phi_2}\big)}{mr^2}-\frac{V_{r\phi_2}}{M_B I_{B,zz}}-2\frac{\dot{r}\dot{\theta}}{r^2}
\end{gather}
\begin{gather}
\frac{\partial \ddot{\phi_2}}{\partial \phi_1}=-\frac{V_{\phi1\phi1}}{mr^2}
\end{gather}
\begin{gather}
\frac{\partial \ddot{\phi_2}}{\partial \phi_2}=-\frac{V_{\phi2\phi2}}{mr^2}-\frac{V_{\phi2\phi2}}{M_B I_{B,zz}}
\end{gather}
\begin{gather}
\frac{\partial \ddot{\phi_2}}{\partial \dot{r}}=2\frac{\dot{\theta}}{r}
\end{gather}
\begin{gather}
\frac{\partial \ddot{\phi_2}}{\partial \dot{\theta}}=2\frac{\dot{r}}{r}
\end{gather}

For second order inertia tensor formulation of mutual potential, partials of the potential are as follows:

\begin{gather}
V_{r}=\frac{GM_AM_B}{r^2}\big(1+\frac{3}{2r^2}\big(I_{A,xx}+I_{A,yy}+I_{A,zz}+I_{B,xx}+I_{B,yy}+I_{B,zz}\\
\nonumber-\frac{3}{2}\big(I_{A,xx}+I_{A,yy}
-cos\big(2\phi_1\big)\big(I_{A,yy}-I_{A,xx}\big)+I_{B,xx}+I_{B,yy}\\
\nonumber-cos\big(2\phi_2\big)\big(I_{B,yy}-I_{B,xx}\big)\big)\big)\big)
\end{gather}
\begin{gather}
V_{\phi_1}=3\frac{GM_AM_B}{2r^3}sin\big(2\phi_1\big)\big(I_{A,yy}-I_{A,xx}\big)
\end{gather}
\begin{gather}
V_{\phi_2}=3\frac{GM_AM_B}{2r^3}sin\big(2\phi_2\big)\big(I_{B,yy}-I_{B,xx}\big)
\end{gather}

\begin{gather}
V_{rr}=2\frac{GM_AM_B}{r^3}-6\frac{GM_AM_B}{r^5}\big(I_{A,xx}+I_{A,yy}+I_{A,zz}+I_{B,xx}+I_{B,yy}+I_{B,zz}\\
\nonumber-\frac{3}{2}\big(I_{A,xx}+I_{A,yy}
-cos\big(2\phi_1\big)\big(I_{A,yy}-I_{A,xx}\big)+I_{B,xx}+I_{B,yy}\\
\nonumber-cos\big(2\phi_2\big)\big(I_{B,yy}-I_{B,xx}\big)\big)\big)
\end{gather}
\begin{gather}
V_{r\phi_1}=-9\frac{GM_AM_B}{2r^4}sin\big(2\phi_1\big)\big(I_{A,yy}-I_{A,xx}\big)
\end{gather}
\begin{gather}
V_{r\phi_2}=-9\frac{GM_AM_B}{2r^4}sin\big(2\phi_2\big)\big(I_{B,yy}-I_{B,xx}\big)
\end{gather}
\begin{gather}
V_{\phi_1\phi_1}=3\frac{GM_AM_B}{r^3}cos\big(2\phi_1\big)\big(I_{A,yy}-I_{A,xx}\big)
\end{gather}
\begin{gather}
V_{\phi_2\phi_2}=3\frac{GM_AM_B}{r^3}cos\big(2\phi_2\big)\big(I_{B,yy}-I_{B,xx}\big)
\end{gather}

\section{Nonplanar Dynamics Matrix}
Within this section we add the notation $\Big(-\Big)^s$ to denote a skew-symmetric matrix operator in addition to the previous tilde notation.
\begin{gather}
\textbf{\textit{A}}=\begin{bmatrix}
0_3&0_3&0_3&0_3&0_3&0_3\\
0_3&\frac{\partial \dot{\vec{\theta}}_1}{\partial \vec{\theta}_1}&0_3&0_3&B_1&0_3\\
0_3&0_3&\frac{\partial \dot{\vec{\theta}}_2}{\partial \vec{\theta}_2}&0_3&0_3&B_2\\
\frac{\partial \ddot{\vec{R}}}{\partial \vec{r}}&0_3&\frac{\partial \ddot{\vec{r}}}{\partial \vec{\theta}_2}&\frac{\partial \ddot{\vec{r}}}{\partial \dot{\vec{r}}}&\frac{\partial \ddot{\vec{r}}}{\partial \vec{\omega}_1}&0_3\\
\frac{\partial \dot{\vec{\omega}}_1}{\partial \vec{r}}&0_3&\frac{\partial \dot{\vec{\omega}}_1}{\partial \vec{\theta}_2}&0_3&\frac{\partial \dot{\vec{\omega}}_1}{\partial \vec{\omega}_1}&0_3\\
\frac{\partial \dot{\vec{\omega}}_2}{\partial \vec{r}}&0_3&\frac{\partial \dot{\vec{\omega}}_2}{\partial \vec{\theta}_2}&0_3&\frac{\partial \dot{\vec{\omega}}_2}{\partial \vec{\omega}_1}&\frac{\partial \dot{\vec{\omega}}_2}{\partial \vec{\omega}_2}
\end{bmatrix}
\end{gather}

\begin{gather}
\frac{\partial \vec{M}_B}{\partial \vec{r}} = -\tilde{\alpha}\frac{\partial^2 U}{\partial \vec{\alpha}\partial \vec{r}}-\tilde{\beta}\frac{\partial^2 U}{\partial \vec{\beta}\partial \vec{r}}-\tilde{\gamma}\frac{\partial^2 U}{\partial \vec{\gamma}\partial \vec{r}}
\end{gather}
\begin{gather}
\frac{\partial \vec{M}_B}{\partial \vec{\theta}_2}=\Big(\frac{\partial U}{\partial \vec{\alpha}}\Big)^s\frac{\partial \vec{\alpha}}{\partial \vec{\theta}_2}+\Big(\frac{\partial U}{\partial \vec{\beta}}\Big)^s\frac{\partial \vec{\beta}}{\partial \vec{\theta}_2}+\Big(\frac{\partial U}{\partial \vec{\gamma}}\Big)^s\frac{\partial \vec{\gamma}}{\partial \vec{\theta}_2}\\
\nonumber-\tilde{\vec{\alpha}}\Big( \tilde{\vec{\alpha}}\frac{\partial^2 U}{\partial \vec{\alpha}^2} +\tilde{\vec{\beta}}\frac{\partial^2 U}{\partial \vec{\alpha}\partial \vec{\beta}}+\tilde{\vec{\gamma}}\frac{\partial^2 U}{\partial \vec{\alpha}\partial \vec{\gamma}}\Big)\\
\nonumber -\tilde{\vec{\beta}}\Big( \tilde{\vec{\alpha}}\frac{\partial^2 U}{\partial \vec{\beta}\partial \vec{\alpha}} +\tilde{\vec{\beta}}\frac{\partial^2 U}{\partial \vec{\beta}^2}+\tilde{\vec{\gamma}}\frac{\partial^2 U}{\partial \vec{\beta}\partial \vec{\gamma}}\Big)\\
\nonumber-\tilde{\vec{\gamma}}\Big( \tilde{\vec{\alpha}}\frac{\partial^2 U}{\partial \vec{\gamma }\partial \vec{\alpha}} +\tilde{\vec{\beta}}\frac{\partial^2 U}{\partial \vec{\gamma}\partial \vec{\beta}}+\tilde{\vec{\gamma}}\frac{\partial^2 U}{\partial \vec{\gamma}^2}\Big)
\end{gather}
\begin{gather}
\frac{\partial \vec{M}_A}{\partial \vec{r}} = -\Big(\frac{\partial U}{\partial \vec{r}}\Big)^s+\tilde{r}\frac{\partial^2 U}{\partial \vec{r}^2}-\frac{\partial \vec{M}_B}{\partial \vec{r}}
\end{gather}
\begin{gather}
\frac{\partial \vec{M}_A}{\partial \vec{\theta}_2}=\tilde{r}\Big( \tilde{\alpha}\frac{\partial^2 U}{\partial \vec{\alpha}\partial \vec{r}}+\tilde{\beta}\frac{\partial^2 U}{\partial \vec{\beta}\partial \vec{r}}+\tilde{\gamma}\frac{\partial^2 U}{\partial \vec{\gamma}\partial \vec{r}}\Big)-\frac{\partial \vec{M}_B}{\partial \vec{\theta}_2}
\end{gather}

\begin{gather}
\frac{\partial \dot{\vec{\theta}}_1}{\partial \vec{\theta}_1}=\frac{\partial B_1}{\partial \vec{\theta}_1}\vec{\omega}_1
\end{gather}
\begin{gather}
\frac{\partial \dot{\vec{\theta}}_2}{\partial \vec{\theta}_2}=\frac{\partial B_2}{\partial \vec{\theta}_2}\vec{\omega}_2
\end{gather}

\begin{gather}
\frac{\partial \ddot{\vec{r}}}{\partial \vec{r}} = \tilde{r}\textbf{\textit{I}}_A^{-1}\frac{\partial \vec{M}_A}{\partial \vec{r}} -\Big(\textbf{\textit{I}}_A^{-1}\Big( \textbf{\textit{I}}_A \vec{\omega}_1\tilde{\omega}_1 +M_A\Big)\Big)^s-\tilde{\omega}_1\tilde{\omega}_1 - \frac{1}{m}\frac{\partial^2 U}{\partial \vec{r}^2}
\end{gather}
\begin{gather}
\frac{\partial \ddot{\vec{r}}}{\partial \vec{\theta}_2} = \tilde{r}\textbf{\textit{I}}_A^{-1} \frac{\partial \vec{M}_A}{\partial \vec{\theta}_2} -\Big( \tilde{\alpha}\frac{\partial^2 U}{\partial \vec{\alpha}\partial \vec{r}}+\tilde{\beta}\frac{\partial^2 U}{\partial \vec{\beta}\partial \vec{r}}+\tilde{\gamma}\frac{\partial^2 U}{\partial \vec{\gamma}\partial \vec{r}}\Big)
\end{gather}
\begin{gather}
\frac{\partial \ddot{\vec{r}}}{\partial \dot{\vec{r}}} = -2\tilde{\omega}_1
\end{gather}
\begin{gather}
\frac{\partial \ddot{\vec{r}}}{\partial \vec{\omega}_1} =  \tilde{r} \textbf{\textit{I}}_A^{-1}\Big( \Big(\textbf{\textit{I}}_A \vec{\omega}_1\Big)^s-\tilde{\omega}_1 \textbf{\textit{I}}_A\Big)+2\dot{\tilde{r}}+\Big(
\tilde{\omega}_1 \vec{r}\Big)^s+\tilde{\omega}_1 \tilde{r}
\end{gather}

\begin{gather}
\frac{\partial \dot{\vec{\omega}}_1}{\partial \vec{r}} = \textbf{\textit{I}}_A^{-1} \frac{\partial \vec{M}_A}{\partial \vec{r}} 
\end{gather}
\begin{gather}
\frac{\partial \dot{\vec{\omega}}_1}{\partial \vec{\theta}_2} = \textbf{\textit{I}}_A^{-1} \frac{\partial \vec{M}_A}{\partial \vec{\theta}_2} 
\end{gather}
\begin{gather}
\frac{\partial \dot{\vec{\omega}}_1}{\partial \vec{\omega}_1} = \textbf{\textit{I}}_A^{-1} \Big( \textbf{\textit{I}}_A \vec{\omega}_1 \tilde{\omega}_1 \textbf{\textit{I}}_A\Big)^s
\end{gather}

\begin{gather}
\frac{\partial \dot{\vec{\omega}}_2}{\partial \vec{r}} = \textbf{\textit{I}}_B^{-1} \frac{\partial \vec{M}_B}{\partial \vec{r}} 
\end{gather}
\begin{gather}
\frac{\partial \dot{\vec{\omega}}_2}{\partial \vec{\theta}_2} = \textbf{\textit{I}}_B \Big( \vec{\omega}_1 +\vec{\omega}_2 \Big)\tilde{\omega}_1 \frac{\partial \textbf{\textit{I}}_B^{-1}}{\partial \vec{\theta}_2} + \textbf{\textit{I}}_B^{-1} \Big( -\tilde{\omega}_1 \Big( \vec{\omega}_1 +\vec{\omega}_2\Big)\frac{\partial \textbf{\textit{I}}_B}{\partial \vec{\theta}_2} + \\
\nonumber \frac{\partial \vec{M}_A}{\partial \vec{\theta}_2} -\Big( \vec{\omega}_1 +\vec{\omega}_2 \Big)\frac{\partial \dot{\textbf{\textit{I}}}_B}{\partial \vec{\theta}_2}\Big) \\
\nonumber+\vec{M}_B \frac{\partial \textbf{\textit{I}}_B^{-1}}{\partial \vec{\theta}_2} -\dot{\textbf{\textit{I}}}_B\Big( \vec{\omega}_1 +\vec{\omega}_2 \Big) \frac{\partial \textbf{\textit{I}}_B^{-1}}{\partial \vec{\theta}_2}-\textbf{\textit{I}}_A^{-1}\frac{\partial \vec{M}_A}{\partial \vec{\theta}_2}
\end{gather}
\begin{gather}
\frac{\partial \dot{\vec{\omega}}_2}{\partial \vec{\omega}_1} = \textbf{\textit{I}}_B^{-1}\Big(\textbf{\textit{I}}_B \Big( \vec{\omega}_1 +\vec{\omega}_2 \Big)^s -\tilde{\omega}_1 \textbf{\textit{I}}_B -\dot{\textbf{\textit{I}}}_B\Big)+\textbf{\textit{I}}_A^{-1} \Big( \tilde{\omega}_1 \textbf{\textit{I}}_A-\Big(\textbf{\textit{I}}_A \vec{\omega}_1\Big)^s\Big)
\end{gather}
\begin{gather}
\frac{\partial \dot{\vec{\omega}}_2}{\partial \vec{\omega}_2} = \textbf{\textit{I}}_B^{-1} \Big(-\tilde{\omega}_1 \textbf{\textit{I}}_B -\dot{\textbf{\textit{I}}}_B + \Big( \vec{\omega}_1 +\vec{\omega}_2 \Big)\frac{\partial \dot{\textbf{\textit{I}}}_B}{\partial \vec{\omega}_2}\Big)
\end{gather}

\subsection{Total Mass Partials of Dynamics Matrix}
Partials of $A$ with respect to $M_T$

\begin{gather}
\frac{\partial^2 \ddot{\vec{r}}}{\partial \vec{r} \partial M_T} = \tilde{r}\frac{\partial \textbf{\textit{I}}_A^{-1}}{\partial M_T}\frac{\partial \vec{M}_A}{\partial \vec{r}}+\textbf{\textit{I}}_A^{-1}\frac{\partial^2 \vec{M}_A}{\partial \vec{r}\partial M_T}\\
\nonumber-\Big(\frac{\partial \textbf{\textit{I}}_A^{-1}}{\partial M_T}\textbf{\textit{I}}_A \vec{\omega}_1 \tilde{\omega}_1+\textbf{\textit{I}}_A^{-1}\frac{\partial \textbf{\textit{I}}_A}{\partial M_T} \vec{\omega}_1 \tilde{\omega}_1 \frac{\partial \textbf{\textit{I}}_A^{-1}}{\partial M_T} \vec{M}_A +\textbf{\textit{I}}_A^{-1}\frac{\partial \vec{M}_A}{\partial M_T}\Big)^s\\
\nonumber+\frac{1}{M_T (\mu-\mu^2)}\frac{\partial^2 U}{\partial \vec{r}^2}-\frac{1}{m}\frac{\partial^3 U}{\partial \vec{r}^2 \partial M_T}
\end{gather}
\begin{gather}
\frac{\partial^2 \ddot{\vec{r}}}{\partial \vec{\theta}_2 \partial M_T} = \tilde{r}\Big(\frac{\partial \textbf{\textit{I}}_A^{-1}}{\partial M_T}\frac{\partial \vec{M}_A}{\partial \vec{\theta}_2}+\textbf{\textit{I}}_A^{-1}\frac{\partial^2 \vec{M}_A}{\partial \vec{\theta}_2 \partial M_T}\Big)\\
\nonumber+\frac{1}{M_T (\mu-\mu^2)}\frac{\partial^2 U}{\partial \vec{r}\partial \vec{\theta}_2}-\frac{1}{m}\frac{\partial^3 U}{\partial \vec{r} \partial \vec{\theta}_2 \partial M_T}
\end{gather}
\begin{gather}
\frac{\partial^2 \ddot{\vec{r}}}{\partial \vec{\omega}_1 \partial M_T} = \tilde{r}\Big(\frac{\partial \textbf{\textit{I}}_A^{-1}}{\partial M_T}\Big(\textbf{\textit{I}}_A \vec{\omega}_1\Big)^s+\textbf{\textit{I}}_A^{-1}\Big(\frac{\partial \textbf{\textit{I}}_A^{-1}}{\partial M_T}\vec{\omega}_1\Big)^s-\frac{\partial \textbf{\textit{I}}_A^{-1}}{\partial M_T}\tilde{\omega}_1 \textbf{\textit{I}}_A \\
\nonumber-\textbf{\textit{I}}_A^{-1}\tilde{\omega}_1\frac{\partial \textbf{\textit{I}}_A}{\partial M_T}\Big)
\end{gather}

\begin{gather}
\frac{\partial^2 \dot{\vec{\omega}}_1}{\partial \vec{r} \partial  M_T} = \frac{\partial \textbf{\textit{I}}_A^{-1}}{\partial M_T}\frac{\partial \vec{M}_A}{\partial\vec{r}}+\textbf{\textit{I}}_A^{-1}\frac{\partial^2 \vec{M}_A}{\partial \vec{r}\partial M_T}
\end{gather}
\begin{gather}
\frac{\partial^2 \dot{\vec{\omega}}_1}{\partial \vec{\theta}_2 \partial  M_T} = \frac{\partial \textbf{\textit{I}}_A^{-1}}{\partial M_T}\frac{\partial \vec{M}_A}{\partial\vec{\theta}_2}+\textbf{\textit{I}}_A^{-1}\frac{\partial^2 \vec{M}_A}{\partial \vec{\theta}_2\partial M_T}
\end{gather}
\begin{gather}
\frac{\partial^2 \dot{\vec{\omega}}_1}{\partial \vec{\omega}_1 \partial  M_T} = \frac{\partial \textbf{\textit{I}}_A^{-1}}{\partial M_T}\Big(\textbf{\textit{I}}_A \vec{\omega}_1\Big)^s+\textbf{\textit{I}}_A^{-1}\Big(\frac{\partial \textbf{\textit{I}}_A}{\partial M_T}\vec{\omega}_1\Big)^s\\
\nonumber-\frac{\partial \textbf{\textit{I}}_A^{-1}}{\partial M_T}\tilde{\omega}_1 \textbf{\textit{I}}_A-\textbf{\textit{I}}_A^{-1}\tilde{\omega}_1\frac{\partial \textbf{\textit{I}}_A}{\partial M_T}
\end{gather}

\begin{gather}
\frac{\partial^2 \dot{\vec{\omega}}_2}{\partial \vec{r} \partial  M_T} = \frac{\partial \textbf{\textit{I}}_B^{-1}}{\partial M_T}\frac{\partial \vec{M}_B}{\partial \vec{r}}+\textbf{\textit{I}}_B^{-1}\frac{\partial^2 \vec{M}_B}{\partial \vec{r}\partial M_T}-\frac{\partial \textbf{\textit{I}}_A^{-1}}{\partial M_T}\frac{\partial \vec{M}_A}{\partial \vec{r}}-\textbf{\textit{I}}_A^{-1}\frac{\partial^2 \vec{M}_A}{\partial \vec{r}\partial M_T}
\end{gather}
\begin{gather}
\frac{\partial^2 \dot{\vec{\omega}}_2}{\partial \vec{\theta}_2 \partial  M_T} = \frac{\partial \textbf{\textit{I}}_B}{\partial M_T}(\vec{\omega}_1+\vec{\omega}_2)\tilde{\omega}_1\frac{\partial \textbf{\textit{I}}_B^{-1}}{\partial \vec{\theta}_2}+\textbf{\textit{I}}_B(\vec{\omega}_1+\vec{\omega}_2)\tilde{\omega}_1\frac{\partial^2 \textbf{\textit{I}}_B^{-1}}{\partial \vec{\theta}_2 \partial M_T}\\
\nonumber-\frac{\partial \textbf{\textit{I}}_B}{\partial M_T}\tilde{\omega}_1(\vec{\omega}_1+\vec{\omega}_2)\frac{\partial^2 \textbf{\textit{I}}_B}{\partial \vec{\theta}_2 \partial M_T}+\frac{\partial \vec{M}_B}{\partial M_T}\frac{\partial \textbf{\textit{I}}_B^{-1}}{\partial \vec{\theta}_2}+\vec{M}_B\frac{\partial^2 \textbf{\textit{I}}_B^{-1}}{\partial \vec{\theta}_2 \partial M_T}\\
\nonumber -\frac{\partial \textbf{\textit{I}}_B^{-1}}{\partial M_T}\frac{\partial \vec{M}_B}{\partial \vec{\theta}_2}-\textbf{\textit{I}}_B^{-1}\frac{\partial^2 \vec{M}_B}{\partial \vec{\theta}_2 \partial M_T}-\frac{\partial \dot{\textbf{\textit{I}}}_B}{\partial M_T}(\vec{\omega}_1+\vec{\omega}_2)\frac{\partial \textbf{\textit{I}}_B^{-1}}{\partial \vec{\theta}_2}\\
\nonumber-\dot{\textbf{\textit{I}}}_B(\vec{\omega}_1+\vec{\omega}_2)\frac{\partial^2 \textbf{\textit{I}}_B^{-1}}{\partial \vec{\theta}_2 \partial M_T}-\frac{\partial \textbf{\textit{I}}_A^{-1}}{\partial M_T}\frac{\partial \vec{M}_A}{\partial \vec{\theta}_2}-\textbf{\textit{I}}_A^{-1}\frac{\partial^2 \vec{M}_A}{\partial \vec{\theta}_2 \partial M_T}\\
\nonumber-\frac{\partial \textbf{\textit{I}}_B^{-1}}{\partial M_T}(\vec{\omega}_1+\vec{\omega}_2)\frac{\partial \dot{\textbf{\textit{I}}}_B}{\partial \vec{\theta}_2}-\textbf{\textit{I}}_B^{-1}(\vec{\omega}_1+\vec{\omega}_2)\frac{\partial^2 \dot{\textbf{\textit{I}}}_B}{\partial \vec{\theta}_2 \partial M_T}
\end{gather}
\begin{gather}
\frac{\partial^2 \dot{\vec{\omega}}_2}{\partial \vec{\omega}_1 \partial  M_T} = \frac{\partial \textbf{\textit{I}}_B^{-1}}{\partial M_T}\Big(\Big(\textbf{\textit{I}}_B (\vec{\omega}_1+\vec{\omega}_2)\Big)^s-\tilde{\omega}_1 \textbf{\textit{I}}_B-\dot{\textbf{\textit{I}}}_B \Big)\\
\nonumber+\textbf{\textit{I}}_B^{-1}\Big( \Big(\frac{\partial \textbf{\textit{I}}_B}{\partial M_T}(\vec{\omega}_1+\vec{\omega}_2)\Big)^s-\tilde{\omega}_1\frac{\partial \textbf{\textit{I}}_B}{\partial M_T}-\frac{\partial \dot{\textbf{\textit{I}}}_B}{\partial M_T}\Big)\\
\nonumber+\frac{\partial \textbf{\textit{I}}_A^{-1}}{\partial M_T}\Big( \tilde{\omega}_1 \textbf{\textit{I}}_A -\Big(\textbf{\textit{I}}_A \vec{\omega}_1\Big)^s\Big)+\textbf{\textit{I}}_A^{-1}\Big( \tilde{\omega}_1 \frac{\partial \textbf{\textit{I}}_A}{\partial M_T}-\Big(\frac{\partial \textbf{\textit{I}}_A}{\partial M_T}\vec{\omega}_1\Big)^s\Big)
\end{gather}
\begin{gather}
\frac{\partial^2 \dot{\vec{\omega}}_2}{\partial \vec{\omega}_2 \partial  M_T} = \frac{\partial \textbf{\textit{I}}_B^{-1}}{\partial M_T}\Big( -\tilde{\omega}_1 \textbf{\textit{I}}_B -\dot{\textbf{\textit{I}}}_B+(\vec{\omega}_1+\vec{\omega}_2)\frac{\partial\dot{\textbf{\textit{I}}}_B}{\partial \vec{\omega}_2}\Big)\\
\nonumber-\textbf{\textit{I}}_B^{-1}\tilde{\omega}_1 \frac{\partial \textbf{\textit{I}}_B}{\partial M_T}-\frac{\partial \dot{\textbf{\textit{I}}}_B}{\partial M_T}+(\vec{\omega}_1+\vec{\omega}_2)\frac{\partial^2\dot{\textbf{\textit{I}}}_B}{\partial \vec{\omega}_2 \partial M_T}
\end{gather}

\subsection{Second Order Inertia Integral Partials of Dynamics Matrix}
Partials of $A$ with respect to $T_i^{jkl}$, where $i$ represents body A or B and $j,k,l=0$ or $2$ with only one index set to $2$, are needed to compute the sensitivity of the eigenvalues to the mass parameters of interest.
\begin{gather}
\frac{\partial^2 \ddot{\vec{r}}}{\partial \vec{r} \partial T_i^{jkl}} = \tilde{r}\Big( \frac{\partial \textbf{\textit{I}}_A^{-1}}{\partial T_i^{jkl} } \frac{\partial \vec{M}_A}{\partial \vec{r}} + \textbf{\textit{I}}_A^{-1}\frac{\partial^2 \vec{M}_A}{\partial \vec{r}\partial T_i^{jkl} }\Big)\\
\nonumber -\Big(\frac{\partial \textbf{\textit{I}}_A^{-1}}{\partial T_i^{jkl} } \textbf{\textit{I}}_A \vec{\omega}_1\tilde{\omega}_1 +\textbf{\textit{I}}_A^{-1} \frac{\partial \textbf{\textit{I}}_A}{\partial T_i^{jkl} } \vec{\omega}_1\tilde{\omega}_1 +\frac{\partial \textbf{\textit{I}}_A^{-1}}{\partial T_i^{jkl} } \vec{M}_A +\textbf{\textit{I}}_A^{-1} \frac{\partial \vec{M}_A}{\partial T_i^{jkl}}\Big)^s\\
\nonumber-\frac{1}{m}\frac{\partial^3 U}{\partial \vec{R}^2 \partial T_i^{jkl}}
\end{gather}
\begin{gather}
\frac{\partial^2 \ddot{\vec{r}}}{\partial \vec{\theta}_2 \partial T_i^{jkl}} = \tilde{r}\Big( \frac{\partial \textbf{\textit{I}}_A^{-1}}{\partial T_i^{jkl}}\frac{\partial \vec{M}_A}{\partial \vec{\theta}_2}+\textbf{\textit{I}}_A^{-1}\frac{\partial^2 \vec{M}_A}{\partial \vec{\theta}_2 \partial T_i^{jkl}}\Big)-\frac{1}{m}\frac{\partial^3 U}{\partial \vec{\theta}_2^2 \partial T_i^{jkl}}
\end{gather}
\begin{gather}
\frac{\partial^2 \ddot{\vec{r}}}{\partial \vec{\omega}_1 \partial T_i^{jkl}} = \tilde{r}\Big(  \frac{\partial \textbf{\textit{I}}_A^{-1}}{\partial T_i^{jkl}}\Big(\textbf{\textit{I}}_A \vec{\omega}_1\Big)^s +\textbf{\textit{I}}_A^{-1} \Big(\frac{\partial \textbf{\textit{I}}_A}{\partial T_i^{jkl}}\vec{\omega}_1\Big)^s\\
\nonumber-\frac{\partial \textbf{\textit{I}}_A^{-1}}{\partial  T_i^{jkl}}\tilde{\omega}_1 \textbf{\textit{I}}_A -\textbf{\textit{I}}_A^{-1}\tilde{\omega}_1 \frac{\partial \textbf{\textit{I}}_A}{\partial T_i^{jkl}}\Big)
\end{gather}

\begin{gather}
\frac{\partial^2 \dot{\vec{\omega}}_1}{\partial \vec{r} \partial  T_i^{jkl}} = \frac{\partial \textbf{\textit{I}}_A^{-1}}{\partial  T_i^{jkl}}\frac{\partial \vec{M}_A}{\partial\vec{r}}+\textbf{\textit{I}}_A^{-1} \frac{\partial^2 \vec{M}_A}{\partial \vec{r}\partial  T_i^{jkl}}
\end{gather}
\begin{gather}
\frac{\partial^2 \dot{\vec{\omega}}_1}{\partial \vec{\theta}_2 \partial  T_i^{jkl}} = \frac{\partial \textbf{\textit{I}}_A^{-1}}{\partial  T_i^{jkl}}\frac{\partial \vec{M}_A}{\partial\vec{\theta}_2}+\textbf{\textit{I}}_A^{-1} \frac{\partial^2 \vec{M}_A}{\partial \vec{\theta}_2\partial  T_i^{jkl}}
\end{gather}
\begin{gather}
\frac{\partial^2 \dot{\vec{\omega}}_1}{\partial \vec{\omega}_1 \partial  T_i^{jkl}} = \frac{\partial \textbf{\textit{I}}_A^{-1}}{\partial T_i^{jkl}}\Big(\textbf{\textit{I}}_A \vec{\omega}_1\Big)^s +\textbf{\textit{I}}_A^{-1}\Big(\frac{\partial \textbf{\textit{I}}_A}{\partial T_i^{jkl}}\vec{\omega}_1\Big)^s\\
\nonumber-\frac{\partial \textbf{\textit{I}}_A^{-1}}{\partial T_i^{jkl}}\tilde{\omega}_1 \textbf{\textit{I}}_A -\textbf{\textit{I}}_A^{-1}\tilde{\omega}_1 \frac{\partial \textbf{\textit{I}}_A}{\partial T_i^{jkl}}
\end{gather}

\begin{gather}
\frac{\partial^2 \dot{\vec{\omega}}_2}{\partial \vec{r} \partial  T_i^{jkl}} = \frac{\partial \textbf{\textit{I}}_B^{-1}}{\partial T_i^{jkl}} \frac{\partial \vec{M}_B}{\partial \vec{r}}+\textbf{\textit{I}}_B^{-1} \frac{\partial^2 \vec{M}_B}{\partial \vec{r} \partial T_i^{jkl}}-\frac{\partial \textbf{\textit{I}}_A^{-1}}{\partial T_i^{jkl}} \frac{\partial \vec{M}_A}{\partial \vec{r}}-\textbf{\textit{I}}_A^{-1} \frac{\partial^2 \vec{M}_A}{\partial \vec{r} \partial T_i^{jkl}}
\end{gather}
\begin{gather}
\frac{\partial^2 \dot{\vec{\omega}}_2}{\partial \vec{\theta}_2 \partial  T_i^{jkl}} = \frac{\partial \textbf{\textit{I}}_B}{\partial T_i^{jkl}} \Big( \vec{\omega}_1 +\vec{\omega}_2\Big)\tilde{\omega}_1\frac{\partial \textbf{\textit{I}}_B^{-1}}{\partial \vec{\theta}_2} + \textbf{\textit{I}}_B \Big( \vec{\omega}_1 +\vec{\omega}_2\Big)\tilde{\omega}_1 \frac{\partial^2 \textbf{\textit{I}}_B^{-1}}{\partial \vec{\theta}_2 \partial T_i^{jkl}}\\
\nonumber -\frac{\partial \textbf{\textit{I}}_B^{-1}}{\partial T_i^{jkl}}\tilde{\omega}_1\Big( \vec{\omega}_1 +\vec{\omega}_2\Big)\frac{\partial \textbf{\textit{I}}_B}{\partial \vec{\theta}_2}-\textbf{\textit{I}}_B^{-1}\tilde{\omega}_1\Big( \vec{\omega}_1 +\vec{\omega}_2\Big)\frac{\partial^2 \textbf{\textit{I}}_B}{\partial \vec{\theta}_2 \partial T_i^{jkl}}\\
\nonumber+\frac{\partial \vec{M}_B}{\partial T_i^{jkl}}\frac{\partial \textbf{\textit{I}}_B^{-1}}{\partial \vec{\theta}_2}+\vec{M}_B\frac{\partial^2 \textbf{\textit{I}}_B^{-1}}{\partial \vec{\theta}_2 \partial T_i^{jkl}}+\frac{\partial \textbf{\textit{I}}_B^{-1}}{\partial T_i^{jkl}}\frac{\partial \vec{M}_B}{\partial \vec{\theta}_2}\\
\nonumber +\textbf{\textit{I}}_B^{-1}\frac{\partial^2 \vec{M}_B}{\partial \vec{\theta}_2 \partial T_i^{jkl}}-\frac{\partial \dot{\textbf{\textit{I}}}_B}{\partial T_i^{jkl}}\Big( \vec{\omega}_1 +\vec{\omega}_2\Big)\frac{\partial \textbf{\textit{I}}_B^{-1}}{\partial \vec{\theta}_2}-\dot{\textbf{\textit{I}}}_B\Big( \vec{\omega}_1 +\vec{\omega}_2\Big)\frac{\partial^2 \textbf{\textit{I}}_B^{-1}}{\partial \vec{\theta}_2 \partial T_i^{jkl}} \\
\nonumber -\frac{\partial \textbf{\textit{I}}_B^{-1}}{\partial T_i^{jkl}}\Big( \vec{\omega}_1 +\vec{\omega}_2\Big)\frac{\partial \dot{\textbf{\textit{I}}}_B}{\partial \vec{\theta}_2}-\textbf{\textit{I}}_B^{-1}\Big( \vec{\omega}_1 +\vec{\omega}_2\Big)\frac{\partial^2 \dot{\textbf{\textit{I}}}_B}{\partial \vec{\theta}_2 \partial T_i^{jkl}}\\
\nonumber-\frac{\partial \textbf{\textit{I}}_A^{-1}}{\partial T_i^{jkl}}\frac{\partial \vec{M}_A}{\partial \vec{\theta}_2}-\textbf{\textit{I}}_A^{-1}\frac{\partial^2 \vec{M}_A}{\partial \vec{\theta}_2 \partial T_i^{jkl}}
\end{gather}
\begin{gather}
\frac{\partial^2 \dot{\vec{\omega}}_2}{\partial \vec{\omega}_1 \partial  T_i^{jkl}} = \frac{\partial \textbf{\textit{I}}_B^{-1}}{\partial T_i^{jkl}}\Big(\textbf{\textit{I}}_B \Big( \vec{\omega}_1 +\vec{\omega}_2\Big)\Big)^s+\textbf{\textit{I}}_B^{-1}\Big(\frac{\partial \textbf{\textit{I}}_B}{\partial T_i^{jkl}}\Big( \vec{\omega}_1 +\vec{\omega}_2\Big)\Big)^s\\
\nonumber-\frac{\partial \textbf{\textit{I}}_B^{-1}}{\partial T_i^{jkl}}\tilde{\omega}_1 \textbf{\textit{I}}_B  -\textbf{\textit{I}}_B^{-1}\tilde{\omega}_1 \frac{\partial \textbf{\textit{I}}_B}{\partial T_i^{jkl}}-\frac{\partial \textbf{\textit{I}}_B^{-1}}{\partial T_i^{jkl}}\dot{\textbf{\textit{I}}}_B-\textbf{\textit{I}}_B^{-1}\frac{\partial \dot{\textbf{\textit{I}}}_B}{\partial T_i^{jkl}}\\
\nonumber -\frac{\partial \textbf{\textit{I}}_A^{-1}}{\partial T_i^{jkl}}\Big(\textbf{\textit{I}}_A \vec{\omega}_1\Big)^s -\textbf{\textit{I}}_A^{-1}\Big(\frac{\partial \textbf{\textit{I}}_A}{\partial T_i^{jkl}}\vec{\omega}_1\Big)^s+\frac{\textbf{\textit{I}}_A^{-1}}{\partial T_i^{jkl}}\tilde{\omega}_1 \textbf{\textit{I}}_A +\textbf{\textit{I}}_A^{-1} \tilde{\omega}_1 \frac{\partial \textbf{\textit{I}}_A}{\partial T_i^{jkl}}
\end{gather}
\begin{gather}
\frac{\partial^2 \dot{\vec{\omega}}_2}{\partial \vec{\omega}_2 \partial  T_i^{jkl}} = -\frac{\partial \textbf{\textit{I}}_B^{-1}}{\partial T_i^{jkl}}\tilde{\omega}_1 \textbf{\textit{I}}_B -\textbf{\textit{I}}_B^{-1}\tilde{\omega}_1 \frac{\partial \textbf{\textit{I}}_B}{\partial T_i^{jkl}}-\frac{\partial \textbf{\textit{I}}_B^{-1}}{\partial T_i^{jkl}}\dot{\textbf{\textit{I}}}_B-\textbf{\textit{I}}_B^{-1}\frac{\partial \dot{\textbf{\textit{I}}}_B}{\partial T_i^{jkl}}\\
\nonumber -\frac{\partial \textbf{\textit{I}}_B^{-1}}{\partial T_i^{jkl}}\Big( \vec{\omega}_1 +\vec{\omega}_2\Big)\frac{\partial \dot{\textbf{\textit{I}}}_B}{\partial \vec{\omega}_2}-\textbf{\textit{I}}_B^{-1}\Big( \vec{\omega}_1 +\vec{\omega}_2\Big)\frac{\partial^2 \dot{\textbf{\textit{I}}}_B}{\partial \vec{\omega}_2 \partial T_i^{jkl}}
\end{gather}

\section{General Use Binary Asteroid Simulator}
As a part of the analysis for this paper we developed a tool for dynamical propagation of binary asteroids of arbitrary shape and expansion order using Eq. 11-13 and the Hou mutual gravity potential described in Appendix A\cite{hou2016mutual}. We have provided the software tool for free use at \href{https://github.com/alex-b-davis/gubas}{https://github.com/alex-b-davis/gubas}. The tool, referred to as the General Use Binary Asteroid Simulator, is intended to provide the planetary science community with an easily used, fast, and high fidelity simulation tool for the numerical integration of binary asteroid dynamics. It does not include the tools for the fundamental frequency analysis performed in this paper. 

The software was designed and implemented to be highly modular to enable a wide set of uses and allow for easy integration into larger tool sets. For this reason the architecture was centered around a C++ executable wrapped in a Python shell. The C++ executable performs the numerical integration and calculation of the inertia integrals while the Python wrapper pre-processes user input from a configuration file to initialize the executable and post-processes the results. This approach allows the user to easily modify the Python shell script to fit their needs. In the standard architecture all interactions are handles through the configuration file and a single command line call to initialize the process. While the software and a detailed user guide can be found by following the github link, Fig. B.11 shows a basic flowchart of the software process.

\begin{figure}[!htb]
    \centering
    \includegraphics[width=.75\textwidth]{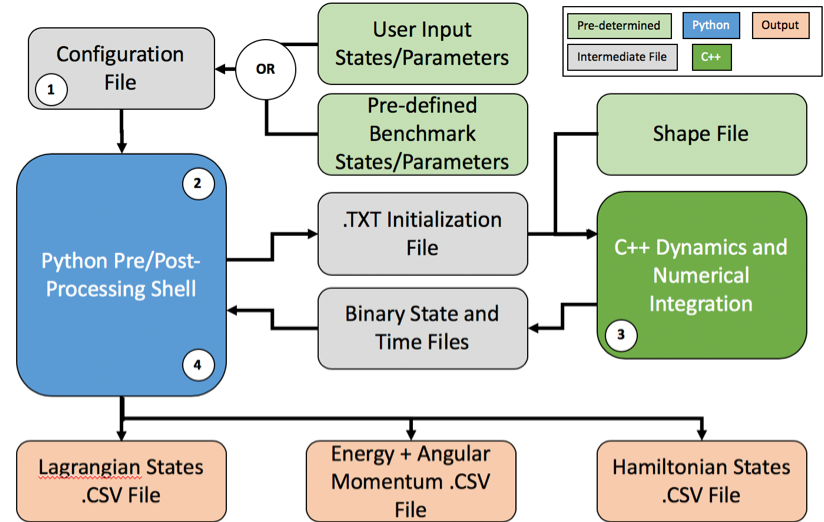}
    \caption{General Use Binary Asteroid Simulator Software Flowchart}
\end{figure}

\bibliographystyle{elsarticle-num} 
\bibliography{references.bib}





\end{document}